# Reconstruction and fast prediction of a 3D flow field based on a variational autoencoder

Gongyan Liu[1], Runze Li[1], Xiaozhou Zhou[2], Tianrui Sun[2], Yufei Zhang[1]*

*1. School of Aerospace Engineering, Tsinghua University, Beijing 100084, China*

*2. Siemens Ltd., Beijing 100084, China*

*Corresponding author: zhangyufei@tsinghua.edu.cn

**ABSTRACT**

Reconstruction and fast prediction of flow fields are important for the improvement of data center operations and energy savings. In this study, an artificial neural network (ANN) and variational autoencoder (VAE) composite model is proposed for the reconstruction and prediction of 3D flowfields with high accuracy and efficiency. The VAE model is trained to extract features of the problem and to realize 3D physical field reconstruction. The ANN is employed to achieve the constructability of the extracted features. A dataset of steady temperature/velocity fields is acquired by computational fluid dynamics and heat transfer (CFD/HT) and fed to train the deep learning model. The proposed ANN-VAE model is experimentally proven to achieve promising field prediction accuracy with a significantly reduced computational cost. Compared to the CFD/HT method, the ANN-VAE method speeds up the physical field prediction by approximately 380,000 times, with mean accuracies of 97.3% for temperature field prediction and 97.9% for velocity field prediction, making it feasible for real-time physical field acquisition.

## I. Introduction

With the rapid development of information technology (IT), data centers have been widely considered an infrastructure of growing importance. Massive amounts of data centers have been built around the world, leading to a significant increase in power consumption **[1]**. Studies on data center energy savings have been considered urgent and profoundly profitable. Usually, data centers are composed of four functional systems, namely, IT equipment, environment equipment, power management equipment and off-site data storage. Each part is indispensable for the overall operation of data centers. The environment system is employed to prevent overheating due to ambient temperature, which is mainly culpable for performance degradation and equipment damage **[2, 3]**. Although it is not the core functional equipment, the environment system consumes approximately 40% of the entire electricity **[4]**. Increasing research is being done to optimize and reduce the computer room air conditioner (CRAC) power consumption rate **[5, 6, 7]**.

Computational fluid dynamics and heat transfer (CFD/HT) simulation is often employed as a powerful tool to predict and analyze the flow-heat transfer coupling problem of runtime data centers **[8, 9, 10]**. ***Liu* et al. (2019) [11]** reported numerous simulation results of internal airflow simulation of server cabinets based on different turbulence models. Full-field prediction of a server room implementing automatic cooling control achieved by CFD/HT was proven to be prominently beneficial for data center air flow design and was expected to be an alternative to a physical monitoring system. However, real-time full-field prediction based on the CFD/HT method requires enormously large computational costs, leading to unattainable resources and time. ***Phan* et al. (2016) [12]** used a reduced order method based on proper orthogonal decomposition (POD) to shorten the simulation time. The proposed POD method is 600 times faster than CFD/HT simulation.

Several studies of data center cooling system control have been carried out based on artificial intelligence (AI). ***De Lorenzi*** **et al. (2012) [13]** proposed an enlightening neural network (NN)-based prediction and control method for air flow in a data center for predicting the local temperature field at the server inlet and generating a control strategy. ***Kansara*** **et al. (2015) [14]** proposed a temperature control model based on neural networks for temperature prediction and control strategy decision-making. ***Lazic*** **et al. (2018) [15]** proposed an application of reinforcement learning to realize data center cooling control, which demonstrated improved operational efficiency compared to existing PID controllers. Some promising attempts at intelligent prediction have also been reported **[16]**. ***Wang*** **et al. (2011) [17]** employed an ANN (artificial neural network) **[18]** to realize real-time prediction of workload thermal effects on data centers. ***Sukthankar*** **et al. (2018) [19]** introduced a prediction method for air temperature in an air-handing unit using ANN. ***Zapate*****r et al. (2016) [20]** proposed a methodology using evolutionary algorithms to make runtime predictions of CPU and inlet temperatures under variable cooling setups, which showed that the average errors are below 2℃ and 0.5℃, respectively.

The development of deep learning (DL) and deep neural network (DNN) techniques has shed new light on physical field prediction, making it possible to achieve real-time 3D physical field prediction. As an important branch of deep learning, generative models enable fast image generation based on a few pieces of critical information. Variational autoencoder (VAE) **[21]** is a fast-developing and widely used generative model that is an unsupervised learning method for complex probabilistic distributions. In 2013, ***Kingma*** **et al. [22]** used probabilities to characterize the latent variable space based on an autoencoder (AE) **[23]**, which substantially improved the generative power of deep learning neural networks. In 2016, ***Doersch*** **et al. [24]** provided a more detailed description and introduction of the structure and principles of VAE. Compared with AE, VAE enables the network to generate more plausible results when generating data in situations that are excluded from the dataset. It is realized by introducing a normal distribution for the construction of hidden variables. The application of VAE is highly prospective in industrial applications. Conventional physical sensors can be replaced by virtual sensors based on VAE, making monitoring systems more intelligent and relatively costless **[25]**.

Scholars further developed various types of derivative networks of VAE. Convolutional VAE is an image generation model developed by ***Sohn*** **et al. [26]** based on the idea of variation and convolutional neural networks (CNNs). Convolutional VAE can also be regarded as a VAE network using a convolutional neural network as an encoder and decoder. CNN is effectual for bidirectional graph-related processing. Based on CNN, ***Liu*** **et al. (2022) [27]** proposed an inversion model for the temperature field of a heat source. ***Zhao*** **et al. (2022) [28]** proposed a physics-informed model for temperature field prediction of heat sources.

Convolutional VAE has received widespread attention and found applications in various industries. In 2021, ***Wang*** **et al. [29]** developed a generative model based on generative deep learning for extracting airfoil flow field features and predicting steady flow fields around supercritical airfoils, within which two-dimensional (2D) pressure field and velocity fields were successfully predicted. The study first extracted representative features of the flow field by a VAE network. Then, an ANN network was used to implement the mapping between the wing shape and the latent variables. Finally, the flow field prediction model for a given wing was achieved by an ANN-VAE composite network. The study demonstrated high accuracy and generality in reconstructing and predicting the flow field around a supercritical airfoil. The ANN-VAE model was also applied to other research areas, such as weather forecasting **[30]** and fluid mechanics studies **[31]**.

In this paper, a prediction method for the temperature and velocity field is proposed by an ANN-VAE model for data center cooling circulations with a fixed geometry and boundary conditions. This method requires the generative model to be trained by a large amount of simulation data. The dataset is acquired through CFD/HT simulations of calculation settings obtained from Latin Hypercubic Sampling (LHS) **[32]**. Network training is based on the backpropagation (BP) algorithm **[33]**. It also requires iterative parametric optimization to acquire a functional surrogate model for the specific problem. In the proposed ANN-VAE model, the function of the ANN part is to compile the input monitoring quantities into feature vectors that VAE can recognize, namely, latent variables. The function of the VAE part is to decode the latent variables into three-dimensional (3D) scalar physical fields. The ResNet **[34]** structure is employed to improve the graphics processing performance of the network. For vector physical fields (such as velocity), the prediction can be achieved through scalar physical field predictions of its components separately.

This paper is organized as follows. In Sec. I, the research background is introduced. In Sec. II, a typical model of a data center is introduced with its cooling cycle. The dataset is acquired based on CFD/HT simulation. Sec. III introduces the structure and function of the AI framework, including the AE feature extraction model, VAE reconstruction model and ANN-VAE prediction model. An evaluation method of the network performance is also introduced. In Sec. IV, the results of the feature extraction and flow field reconstruction are presented. Then, the ANN-VAE network performance of temperature and velocity field prediction is studied. In Sec. V, a summary of the research and an outlook on future research are provided.

## II. Data center cooling cycle

### A. Configuration and parameters of the data center cooling cycle

A typical data center consists of four spatial components: equipment room, air conditioning room, cold air supply layer, and hot air return layer, as shown in **Fig. 1.** The data center in this paper is arranged in a rectangular closed room 16 m long, 14 m wide and 5 m high. A detailed schematic figure of the cooling cycle is shown in **Fig. 2**.

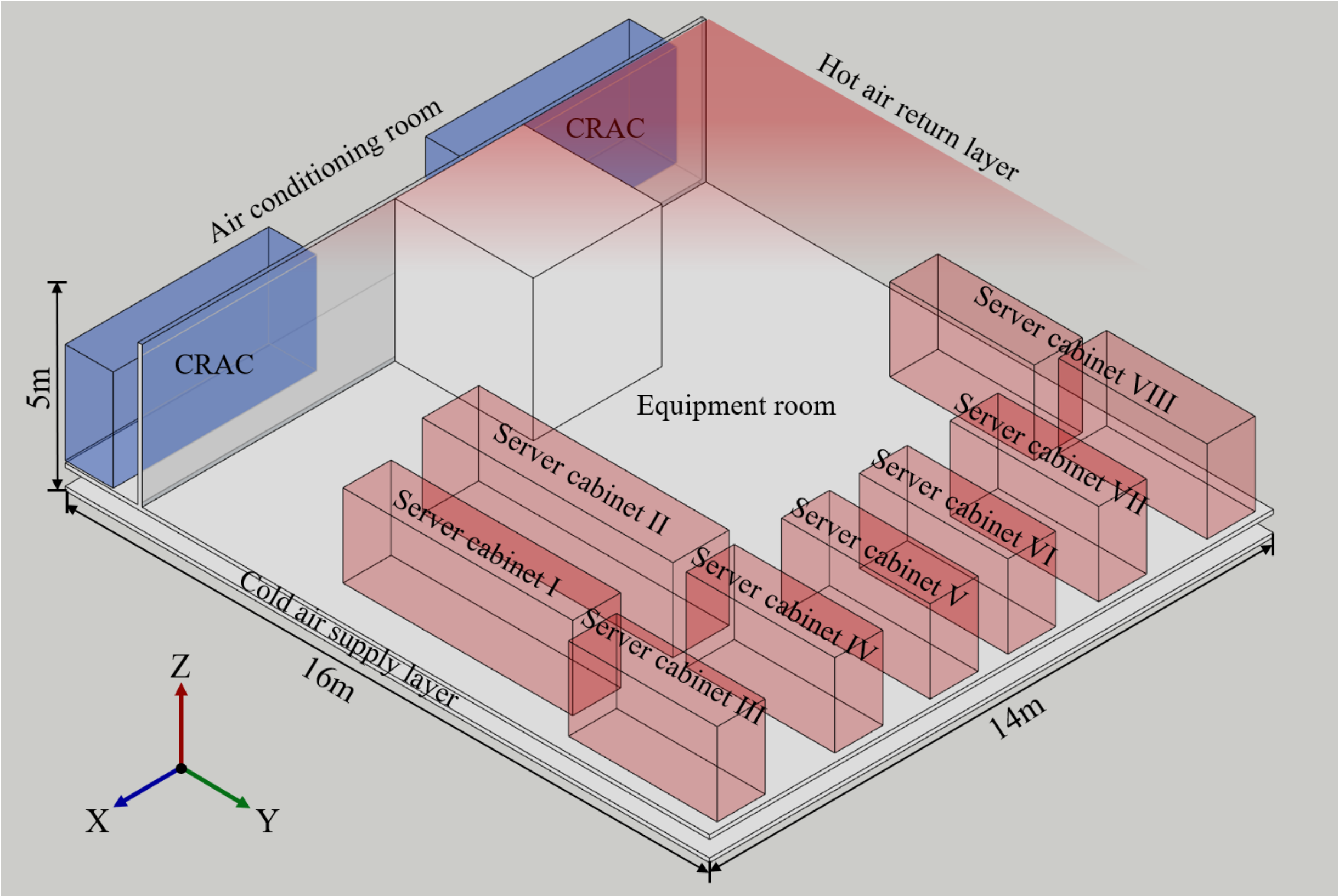

**Fig. 1** Arrangement of the data center

(1) Equipment room. Designed to accommodate IT equipment, generating heat sustainably when operating, causing a significant rise in local temperature and a risk of equipment burnout. As shown in **Fig. 1**, the IT equipment in the present data center is grouped into 8 server cabinets.

(2) Air conditioning room**.** Designed to accommodate 2 CRAC units and to drive the data center cooling cycle.

(3) Cold air supply layer and hot air return layer. An air flow path is designed to improve the

cooling efficiency of the equipment. Cold air is supplied from the floor of the equipment room. The hot air is collected at the top. This cooling cycle enhances the adequacy of convection heat transfer by extending the mixing time for heated and cold air.

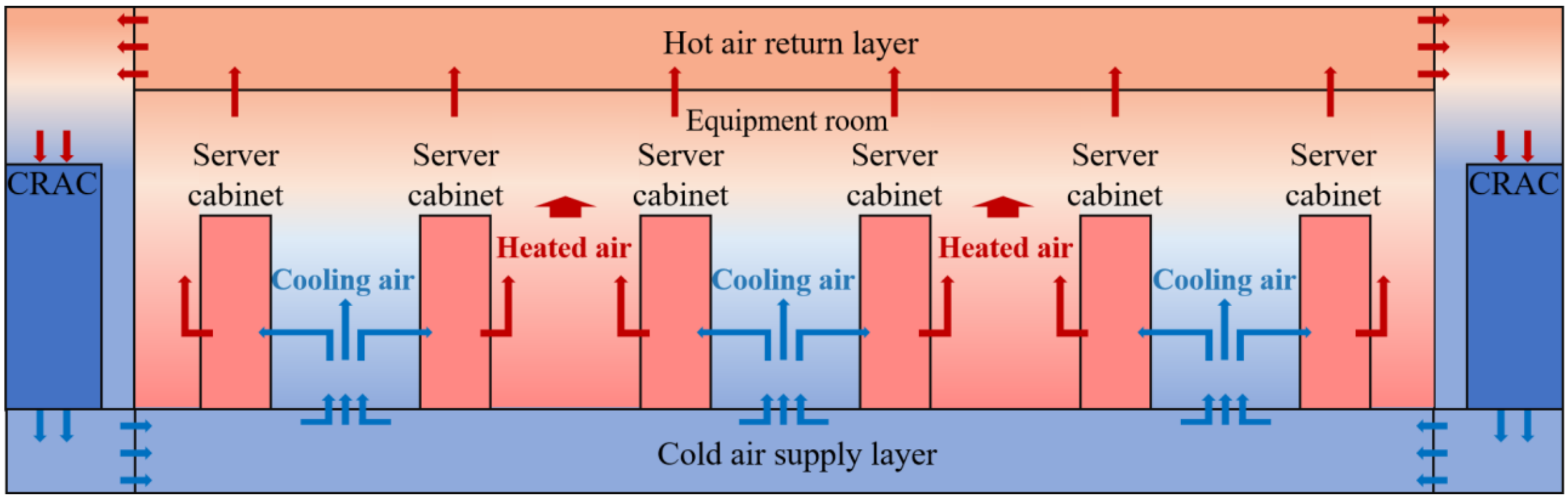

**Fig. 2** Air cooling cycle of the present data center

Because the geometry of the data center is fixed, the flow fields are completely determined by the operating status of the CRACs and server cabinets, which is further interpreted in terms of 10 independent parameters in **Table 1**. The ranges of the parameters are also listed in **Table 1**. When the data center is operating under different conditions, CFD/HT is applied to obtain the flow field and temperature distribution.

**Table 1** Operating parameters of the CRAC and server cabinet

| Type | Parameter | Range of values |
|---|---|---|
| CRAC | Air supply temperature | 20~24℃ |
| | Air supply rate | 3~6 $m^3$ /s |
| Server power consumption | Cabinet I | 0~2000 W |
| | Cabinet II | 0~2000 W |
| | Cabinet III | 0~2000 W |
| | Cabinet IV | 0~2000 W |
| | Cabinet V | 0~2000 W |
| | Cabinet VI | 0~2000 W |
| | Cabinet VII | 0~2000 W |
| | Cabinet VIII | 0~2000 W |

## B. Cooling cycle simulation based on the CFD/HT method

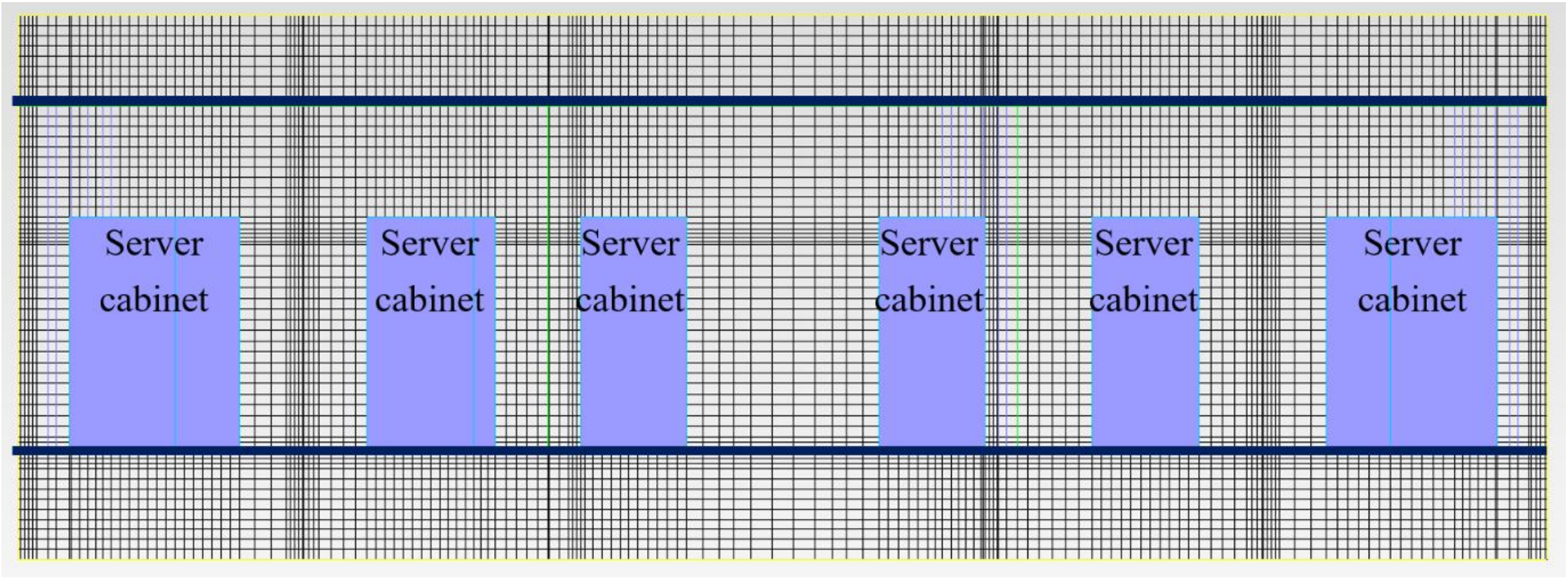

**Fig. 3** Computational domain and grid

The variations in the temperature and velocity fields are of great concern during the cooling cycle. In this study, a CFD/HT method is applied by the commercial code Siemens Flovent, which realizes a numerical solution of steady-state air flow. The computational grid is shown in **Fig. 3**, which contains 1,076,820 cells in total. The mesh grid is refined at the inlet and outlet areas of the CRACs and server cabinets to capture the large temperature gradient. **Fig. 4** presents a sample of the temperature field, which is demonstrated by several planes sliced from the room. It can be seen that the temperature of most of the data center is between 20℃ and 35℃. The XY-planes at Z=1.0 m and 2.4 m, which are planes sliced from the horizontal direction, show that the high-temperature region is near the server cabinets. The temperature distribution in the vertical direction can be seen from the XZ- and YZ-planes. The temperature is low near the floor and is high near the roof.

The computation grid has more than 1 million grid cells, which requires a large amount of memory for AI model training. Twenty layers of the flowfield are extracted from the CFD/HT result to reduce the training cost of the problem. Each layer has 137×131 grid points, so the total grid number of the extracted flow field is 358,940 (i.e., 137×131×20) points.

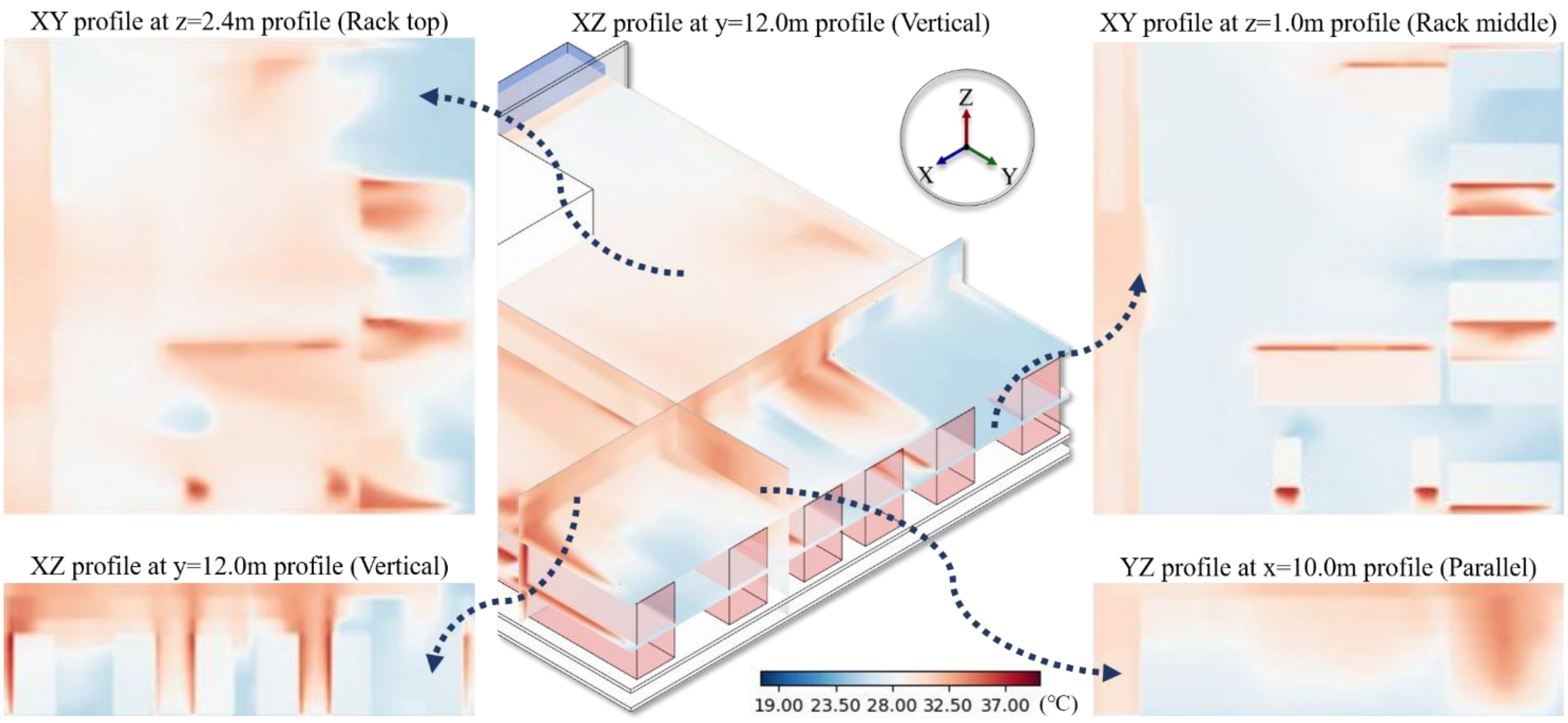

**Fig. 4** Sample dataset (temperature field)

### C. Dataset acquisition of the data center cooling

The model training of the deep learning method requires a large number of samples of the data center cooling cycle. In the present study, both scalar temperature and vector velocity fields are expected to be trained and predicted quickly. The parameter ranges of the cooling cycle are listed in **Table 1**. The dataset for AI model training and testing should cover various operating parameters to ensure the comprehensiveness of the dataset. Consequently, a two-level LHS method is proposed, as shown in **Fig. 5**. The CRAC cooling parameters are sampled at level 1, and the power consumptions of the cabinets are sampled at level 2. At level 1, the cooling temperature and cold air supply rate of the CRACs are sampled by LHS with a size of 10, marking 10 different CRAC cooling states. At level 2, the operation power consumptions of the 8 server cabinets are sampled by LHS with a size of 500, marking 500 different heating conditions of the equipment room. Then, the heating conditions are matched separately to the 10 CRAC cooling states, denoted as 10 subsets. Consequently, the total number of datasets is 5,000 (10×500). All of the samples are computed by the CFD/HT method.

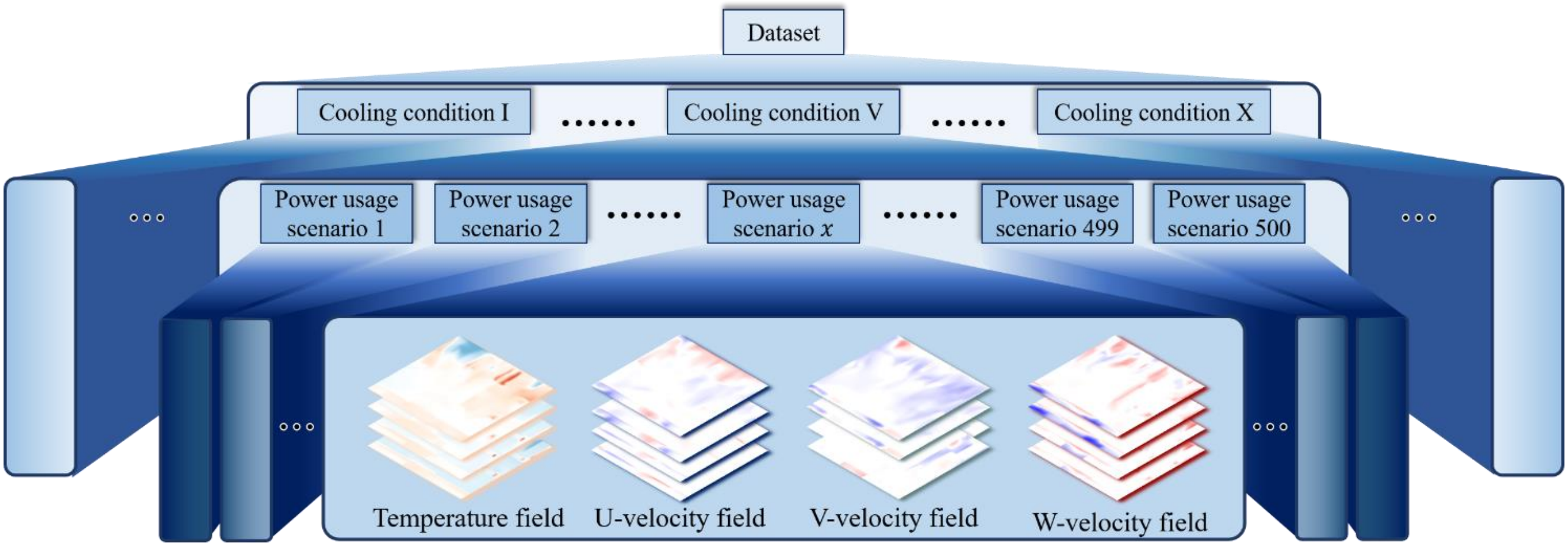

**Fig. 5** Dataset structure sketch of the two-level LHS sampling

The dataset is divided into three parts: the training set, the validation set and the test set, as shown in **Fig. 6**. Eighty-one percent of the samples are set as the training set to adjust the ANN-VAE model parameters. Nine percent of the samples are in the validation set to check the performance of the model during the training process and to provide a performance reference for tuning optimization. Ten percent of the samples are in the test set for the final test of the model performance, i.e., determining whether the model can achieve the prediction goals.

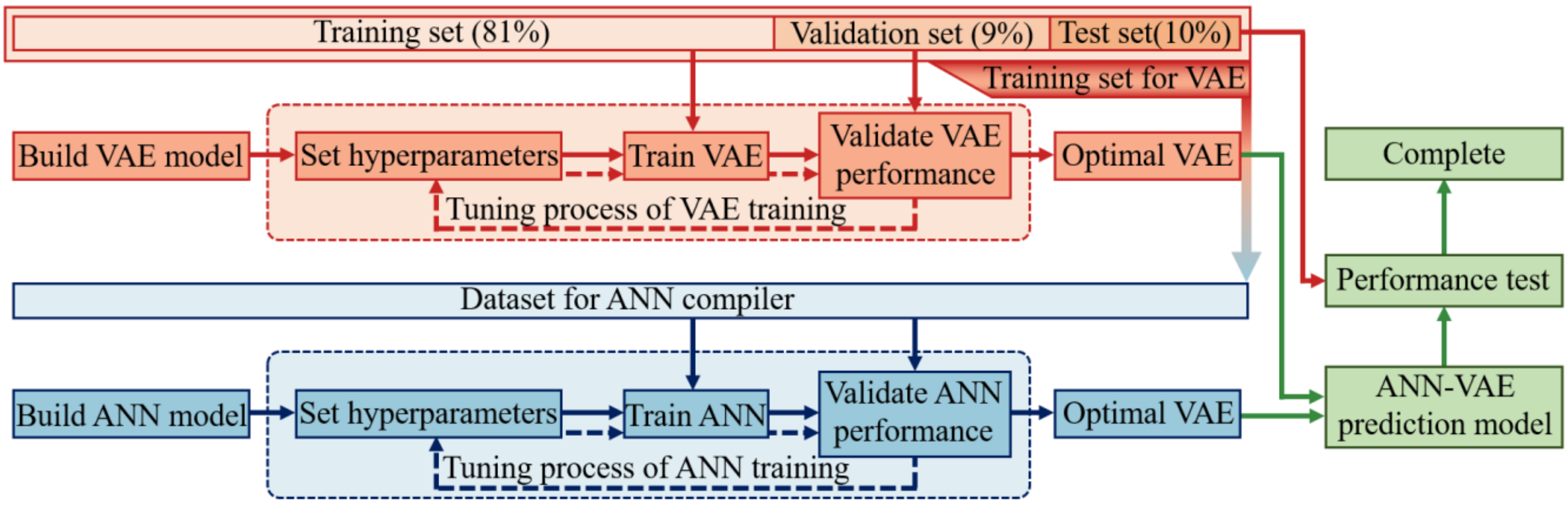

**Fig. 6** Sketch of the dataset splitting and training process

## III. Framework introduction and training implementation

In this paper, an ANN-VAE composite network is proposed to realize fast reconstruction and prediction of flow fields. A diagrammatic sketch of the model and its training process is shown in **Fig. 7**. First, an AE framework is established and trained to obtain the CNN structure for constructing the VAE model. Then, a VAE model is built by introducing the variational principle to its latent space based on the AE model. The VAE is used to accurately reconstruct the flow fields. Afterward, an ANN model is established and trained to connect the input parameters to the latent space of VAE. Finally, the trained ANN is connected to the decoder of the VAE, realizing fast predictions of flow fields.

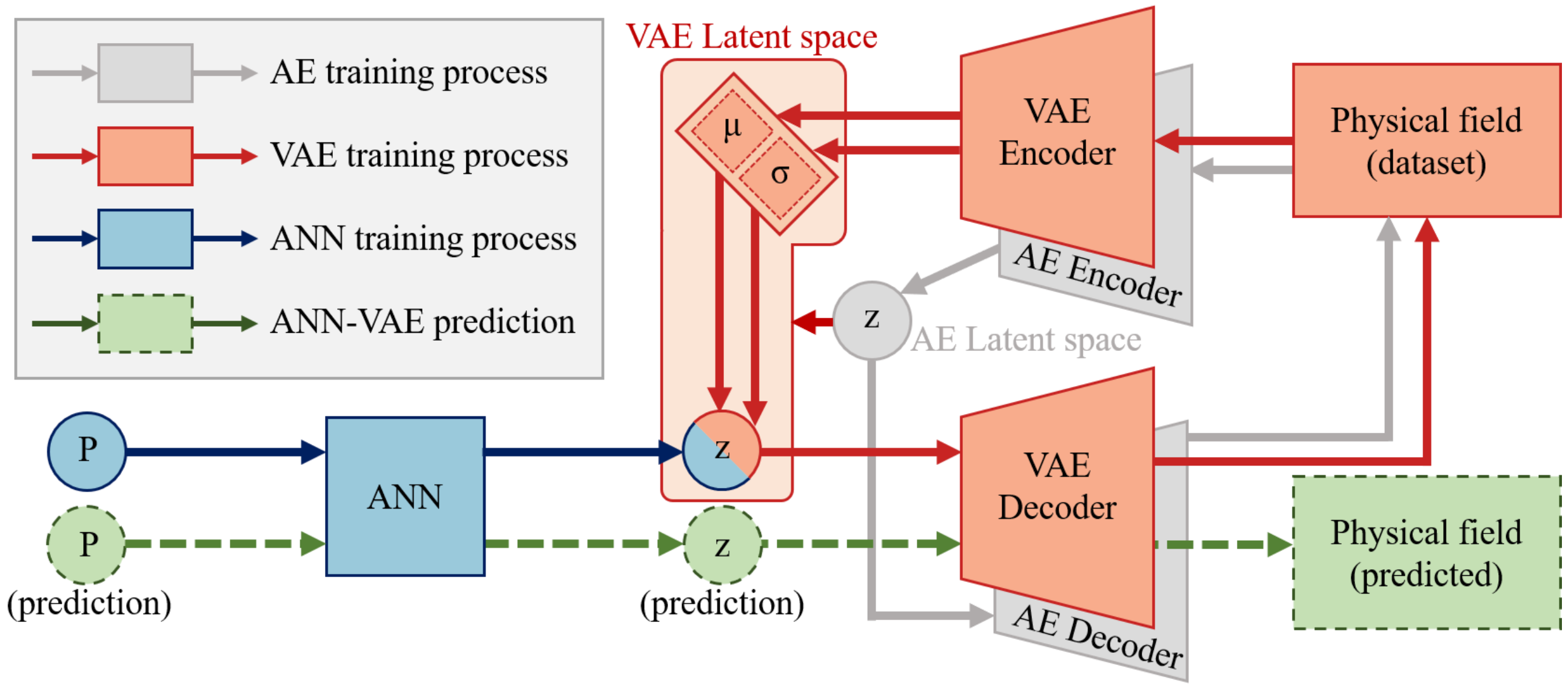

**Fig. 7** Sketch of the ANN-VAE model framework

### A. Model framework of feature extraction based on AE

Directly tuning the VAE is complex because the process contains both structure design and hyperparametric optimization. Consequently, an AE model is initially applied to explore and verify the basic CNN structure configuration for flow field reconstruction. Both AE and VAE consist of five sections: input, encoder, latent vector, decoder and output. The structures of the encoder and decoder and the dimension of the latent space are tested in the AE model. Then, the verified structure and dimension of the latent space are used to construct the VAE model.

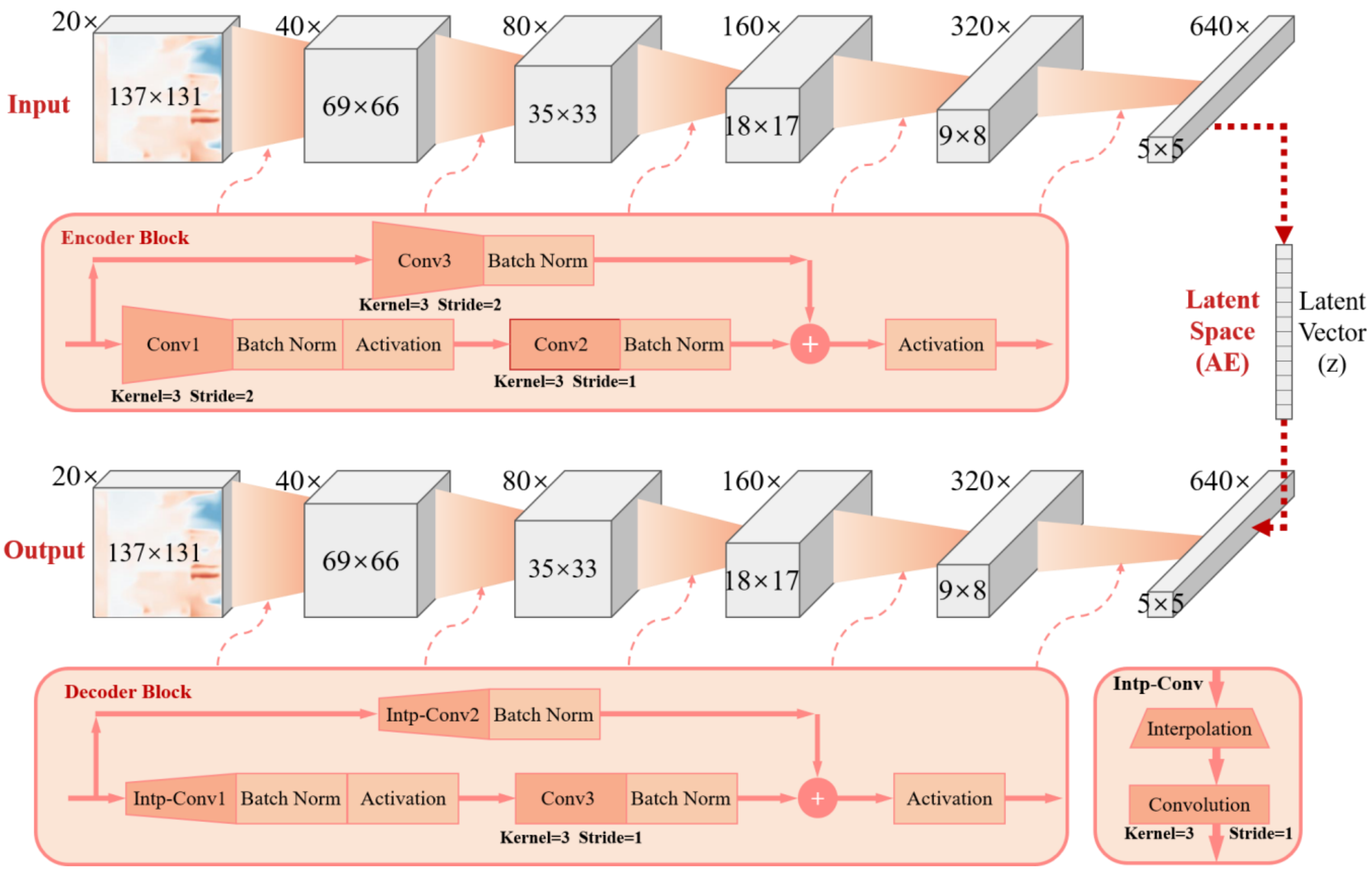

**Fig. 8** The detailed structure of the AE based on the ResNet encoder and decoder

The detailed structure of the AE is shown in **Fig. 8**. The objective of the encoder is to realize reversible data compression for the input flow fields. The encoded data are fed to a latent vector space by a fully connected layer. The dimension of the latent vector space determines the feature number of effectively expressing the physical problem. In the last part of the AE model, the latent vector is decoded to a flow field of the same size as the input data.

In this study, the encoder and decoder are established based on the ResNet structure **[34]**. ResNet is a newly developed variant of CNN. The import of the residual shortcut enables a deeper network structure. The input of the encoder is a scalar matrix of $137\times131\times20$ extracted from a flow field. Then, the input data are sequentially processed by 5 encoder blocks, with each block containing 3 convolution layers, as shown in **Fig. 8**. The convolution stride of Conv1 and Conv3 is 2 in each encoder block, serving as a compressor unit. The input data are finally compressed to size $5\times5\times640$ by the encoder.

The latent vector is designed to be an artificial data bottleneck. The dimension of the latent vector is a decisive structural parameter for AE and VAE. In this study, 8 AE networks with different latent dimensions are trained and compared. The smallest latent dimension is 10, which is equal to the number of operation parameters. A lower dimension might be mathematically unreasonable to maintain the complexity of the flow fields. The other 7 AE networks have latent dimensions ranging from $2^4$ to $2^{10}$. All 8 networks are evaluated by the validation set after training, and an optimal latent dimension is obtained. The optimal dimension of the AE latent space is later adopted for the VAE latent space. The results are presented in Section IV. Denote the encoding mapping as $f_{\mathrm{EC}}$ and the decoding mapping as $f_{\mathrm{DC}}$. The overall mathematical process of AE can be expressed as Eq. (1) and (2), where $x$ is the input, $z$ is the latent vector and $x^{'}$ is the output. The training process is a mapping fitting process for $f_{\mathrm{EC}}$ and $f_{\mathrm{DC}}$, aiming to minimize the difference between $x'$ and $x$.

$$f_{\mathrm{EC}}(Wx+b)=z \tag{1}$$

$$f_{\mathrm{DC}}(z)=x' \tag{2}$$

During the training process of AE, the difference between $x'$ and $x$ is quantified by a loss function. In this study, mean square error (MSE) is adopted as the loss function. As in Eq. (3), MSE is applied to evaluate the difference between the reconstructed field and the input field.

$$loss_{AE}=MSE=\frac{\sum_{i=1}^{n}(x_i-\widehat{x_i})^2}{n} \tag{3}$$

After AE training, the reconstruction performance of the AE is evaluated and analyzed by the mean absolute error (MAE) of the reconstructed and input data, which is more visual for error analysis of flow fields.

$$MAE = \frac{\sum_{i=1}^{n}|x_i - \widehat{x_i}|}{n} \tag{4}$$

## B. Model framework of flow field reconstruction based on VAE

Although AE is capable of data reconstruction, its generalization performance is relatively inadequate due to the simplicity of the latent space structure. Common training failures of AE include overfitting **[35]**, gradient explosion **[36]**, data bias **[37]**, etc. The reconstruction process of the AE model is highly data dependent, leading to a lack of generation accuracy.

**Table 2** Latent space of AE and VAE

| VAE | Layer | Layer input | Layer output | Data amount |
|---|---|---|---|---|
| Latent space (AE) | Fully connected layer 1 | 5×5×640 | $z$ vector | Latent dimension |
| | Latent vector ($z$) | Latent dimension=10, 16, 32, 64, 128, 256, 512, 1024 | | |
| | Fully connected layer 2 | $z$ vector | 5×5×640 | 16,000 |
| Latent space (VAE) | Fully connected layer 1 ($\mu$) | 5×5×640 | $\mu$ vector | Latent dimension |
| | Fully connected layer 2 ($\sigma$) | 5×5×640 | $\sigma$ vector | Latent dimension |
| | Reparameterize layer | $\mu$ and $\sigma$ vector | $z$ vector | Latent dimension |
| | Latent vector ($z$) | Latent dimension: determined from feature extraction by AE | | |
| | Fully connected layer 3 | $z$ vector | 5×5×640 | 16,000 |

Accordingly, the variational principle **[38]** was introduced to the latent space, leading to the development of VAE. As shown in **Fig. 9**, VAE shares the same encoder and decoder frameworks as AE, and the main difference lies in the latent space, as shown in **Table 2**. In VAE, the encoded result is connected by 2 separate fully connected layers to a $\mu$ vector and a $\sigma$ vector. The $\mu$ and $\sigma$ vectors correspond to the mean value and the standard deviation. Then, a latent vector is sampled from a set of normal distributions defined by the mean value $\mu_i$ and standard deviation $\sigma_i$, in a process called reparameterization. The main advantage of VAE is that the latent space is artificially constrained by the variational principle, regulating the distribution of latent variables.

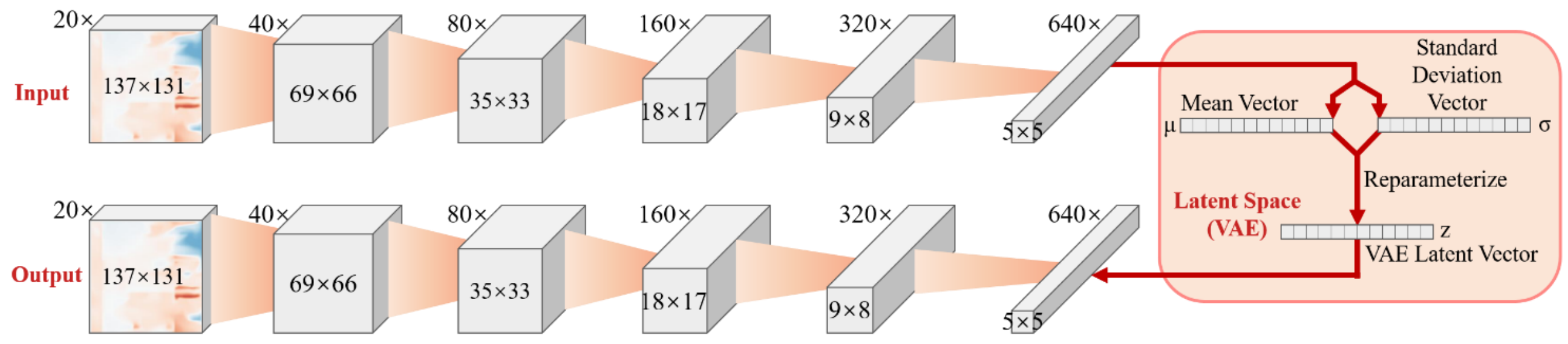

**Fig. 9** The detailed structure of VAE

$$loss_{VAE} = MSE + weight \times KLD \tag{5}$$

The loss function of VAE is composed of two parts, as shown in **Eq. (5)**. The MSE part is applied for difference evaluation between the output and input. The Kullback–Leibler divergence (KLD) **[39]** part is used to evaluate the difference between the distribution in the model network and the standard normal distribution. The expression of KLD is shown in Eq. (6) for multidimensional normal distribution spaces.

$$KL(p||q) = \frac{1}{2}\left[\log\frac{\det(\Sigma_2)}{\det(\Sigma_1)} - d + tr(\Sigma_2^{-1}\Sigma_1) + (\mu_2 - \mu_1)^T\Sigma_2^{-1}(\mu_2 - \mu_1)\right] \tag{6}$$

The weight of KLD determines the constraint strength for the latent space. A large weight of KLD in the loss function will make the model concentrate the latent variable distribution. However, an overweighed KLD may lead to detail blindness of the reconstruction. An inadequate weight of KLD may result in an insufficient constraint for the latent space. In this study, the VAE model with KLD weights ranging from $10^{-3}$ to $10^{-8}$ is trained and tested. The reconstruction performance will be presented in Section IV.

### C. Flowfield prediction based on ANN-VAE

An ANN-VAE deep learning network is constructed and trained to achieve 3D flow field prediction. The composite model is proposed to surrogate the CFD/HT simulation under specific geometric and physical conditions. The ANN-VAE model is formed by combining an ANN compiler and a VAE decoder. The VAE decoder is utilized directly from the optimal VAE model in Section III.B. It is used to decode the latent vector $z$ into a 3D physical field, i.e., the output layer.

The ANN compiler is used to map the input operation parameter $p$ into the latent vector $z$ that can be recognized by the VAE decoder. The ANN structure is shown in **Fig. 10**. The input layer is a 10-dimensional operation parameter vector. Three fully connected hidden layers are set next to the input layer. The node size of each hidden layer is 2048, ensuring sufficient network complexity. The output layer is a vector of the same size as the VAE latent space.

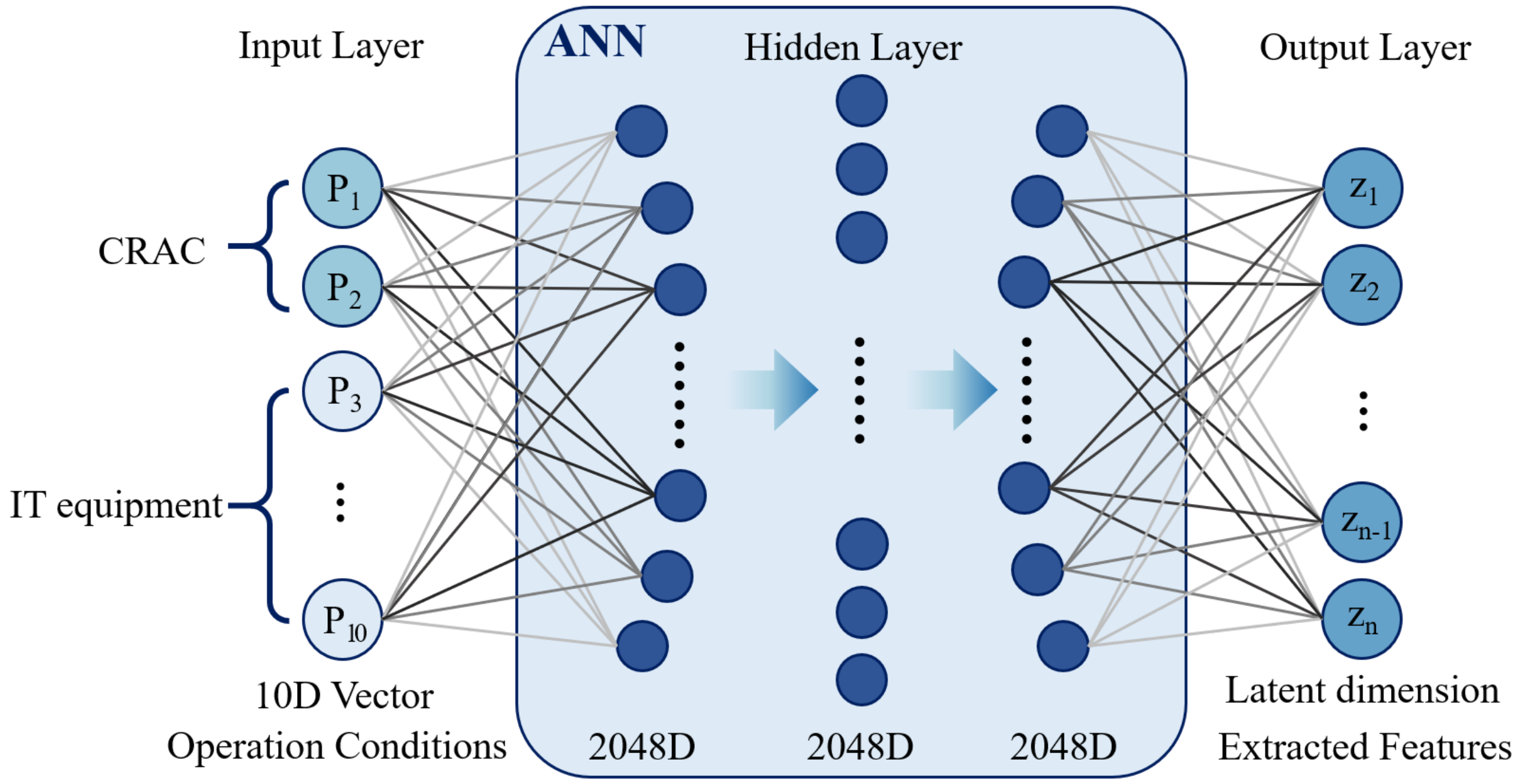

**Fig. 10** The detailed structure of the ANN model

The training process of ANN is driven by a dataset generated by the VAE. The dataset contains 5,000 sets of $p-z$ mapping relations. Each $p$ and $z$ corresponds to a simulation sample, where $z$ is acquired by encoding the flow field data. The ANN and the decoder of VAE are connected together after the ANN is properly trained. Finally, the ANN-VAE model realizes fast prediction of the flow field with an input of 10 operation parameters, as shown in **Fig. 11**.

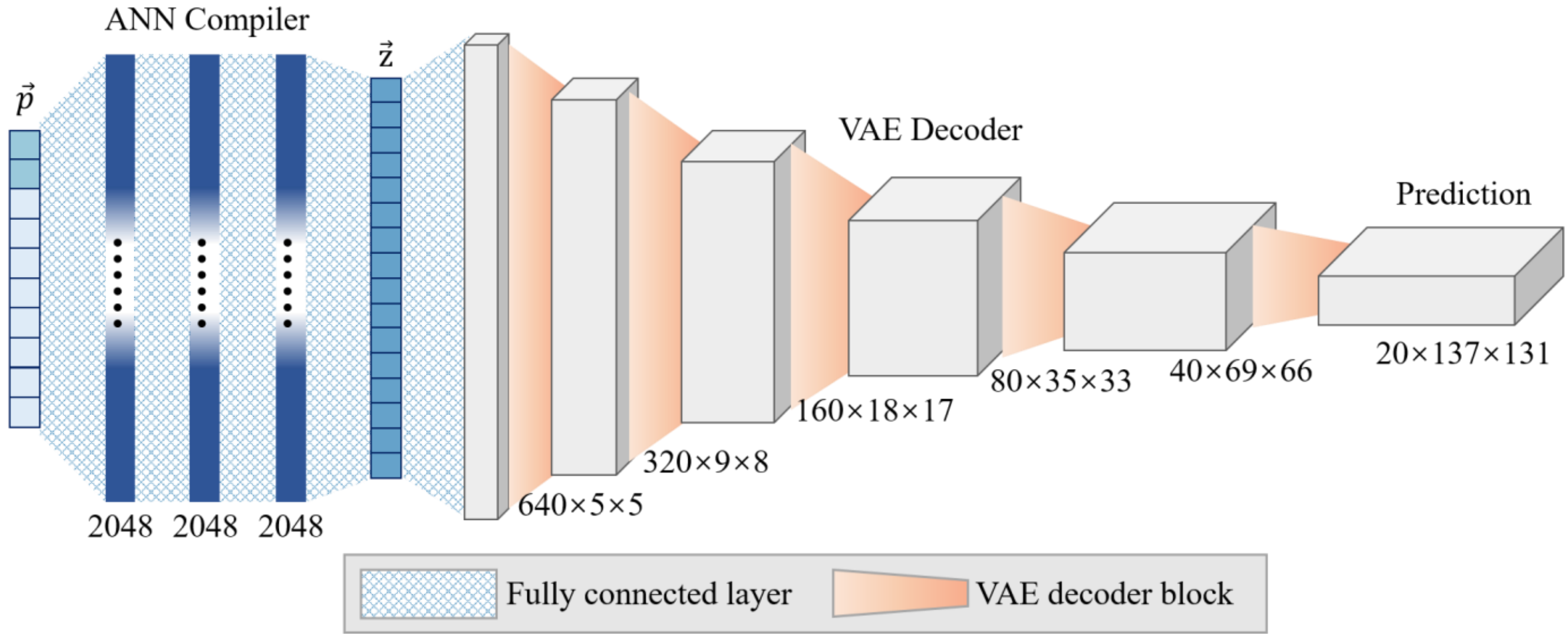

**Fig. 11 Structure of the** ANN-VAE prediction model

MAE is employed to illustrate the reconstruction and generation performance. Additionally, the absolute error distribution over the reconstruction domain is computed. The acceptable absolute error is 1℃ for the temperature field, while the acceptable absolute error is $0.1 \times \sqrt{3}$m/s for the velocity field. The overall evaluation is based on the test set, which was previously untrained by the ANN-VAE network.

## IV. RESULTS AND DISCUSSION

In this section, detailed results of the model training and testing of the composite ANN-VAE model are presented. First, the AE-based feature extraction and VAE-based flow field reconstruction with the temperature field as an example are tested. Then, the performance of the ANN-VAE model for scalar temperature prediction is evaluated. Finally, the prediction accuracy of the velocity field is tested by the ANN-VAE model.

### A. Feature extraction and reconstruction performance

The process of selecting the dimension of the latent variable is implemented using the AE model. The training target of the AE is the reconstruction accuracy of the temperature field. Eight latent dimension numbers ranging from 10 to 1024 are trained and tested. The convergence histories of the loss function with the training epochs are shown in **Fig. 12**. At epoch 300, almost all of the models have reached a relative convergence, with the declines of their loss functions significantly slowing down.

Intuitively, the AE model performs more robustly with a larger latent dimension because it allows greater diversity of feature extraction, which might be beneficial for field reconstruction. However, the mean square error (MSE) shown in **Fig. 12** indicates that a large latent dimension also increases the difficulty of training, which is illustrated by the convergence speed of the loss function. Additionally, the mean absolute error (MAE) of the temperature field of the validation dataset is calculated and compared in **Fig. 13**, which has a similar tendency as the MSE.

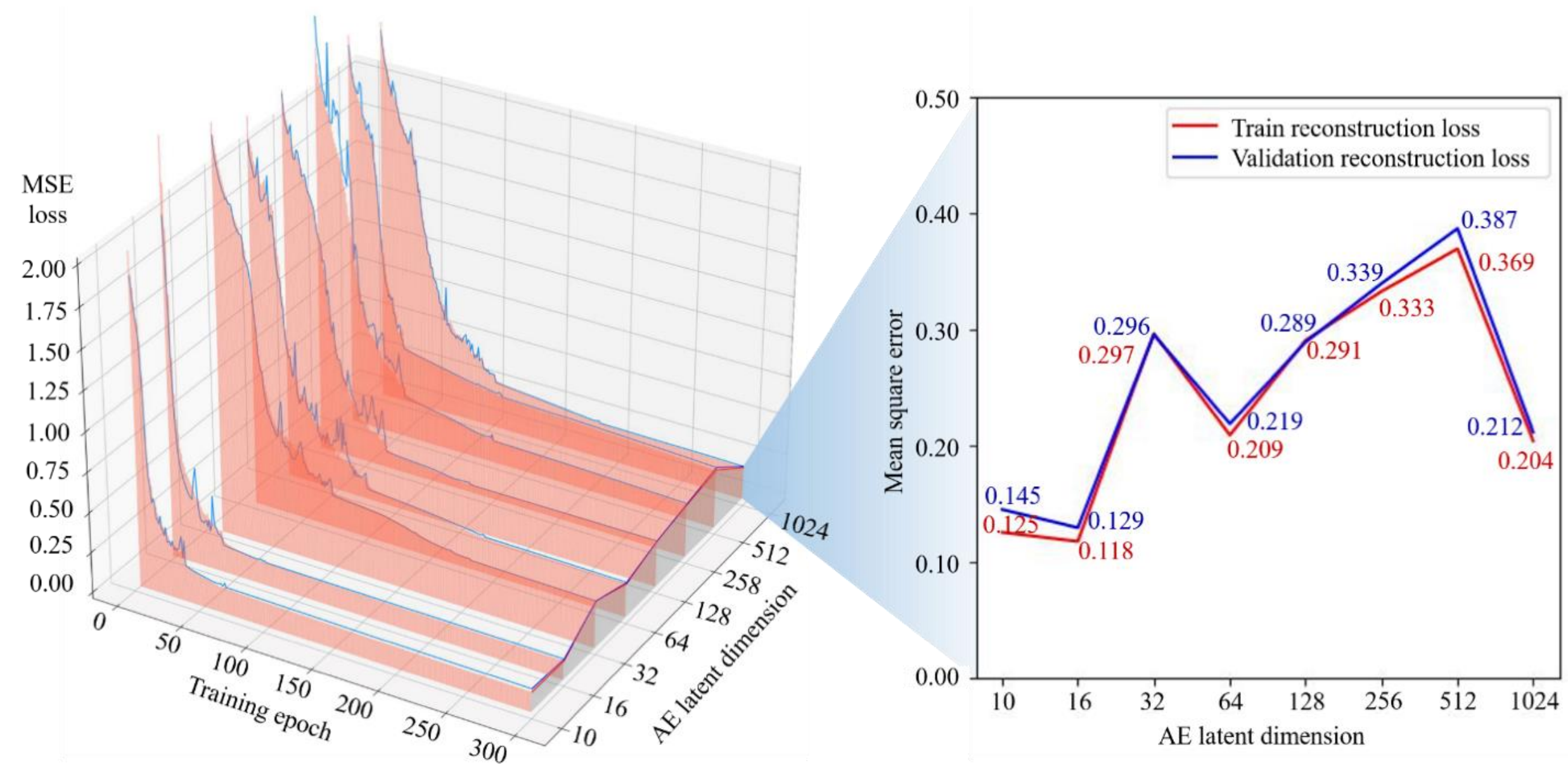

**Fig. 12** Loss convergence of the AE training process with different latent dimensions

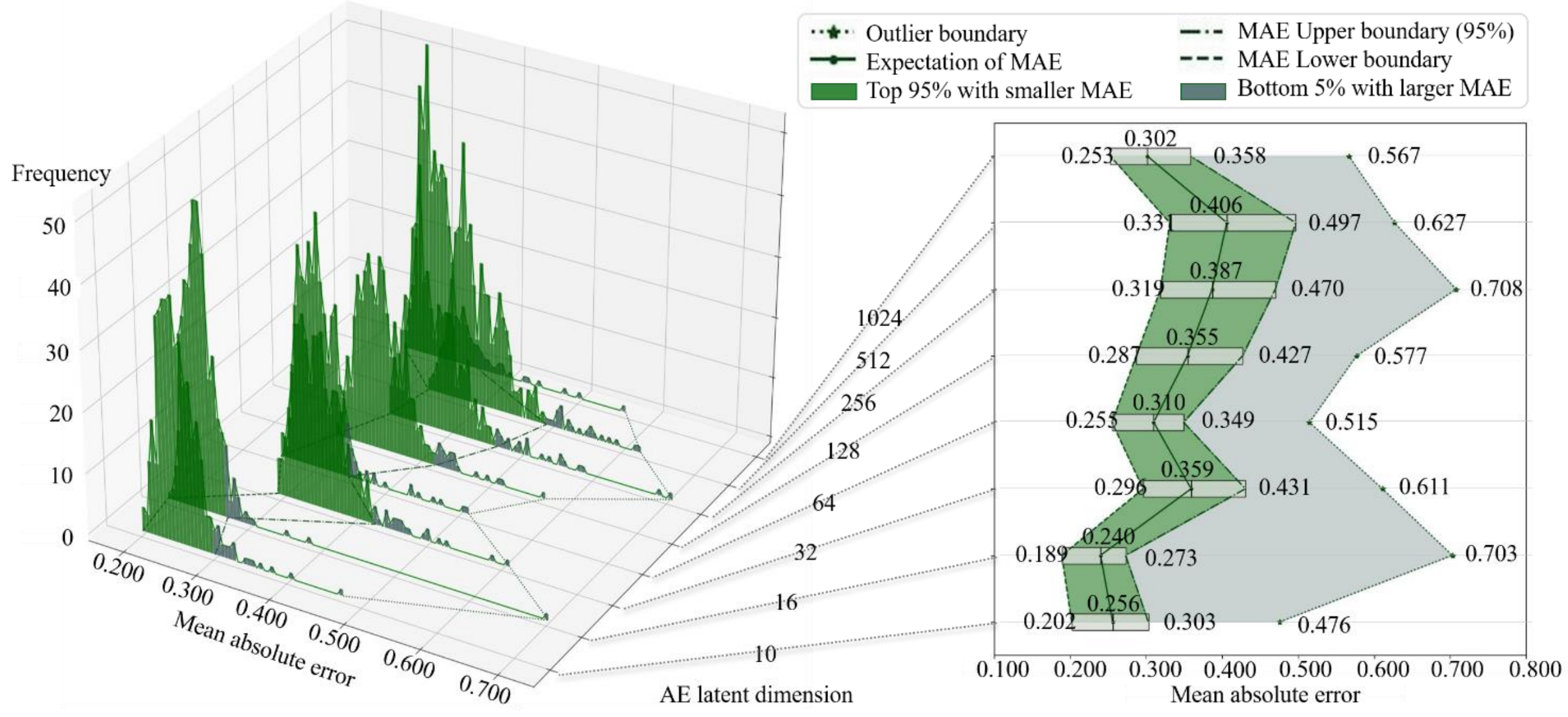

**Fig. 13** Statistics and analysis of the mean absolute error of AE reconstruction

As shown in **Fig. 13**, the AE model with a latent dimension number of 16 delivers an optimal performance among the 8 latent dimension numbers. With this latent dimension number, the reconstruction accuracy is maximized. Ninety-five percent of the samples in the validation set have an MAE less than 0.273℃, and the minimum MAE is as low as 0.240℃. However, approximately 1% of the reconstructions are relatively inaccurate. The maximum MAE of reconstruction is 0.703℃. This is due to overfitting caused by the inherent flaws of the AE latent space.

A comparison of feature extraction performance for a temperature field is illustrated based on the trained AE models. A random field sample is reconstructed by the AE models with different latent dimensions. The temperature distribution located at the waist of the server cabinets (Z=1.0 m, XY profile) is extracted and compared. The reconstruction results are shown in **Fig. 14**. The temperature fields are reconstructed properly for all 8 latent dimensions. The large error regions are generally concentrated at the cabinet outlet. The comparison of the error distribution also shows that the 16-dimensional latent space is an optimal choice.

In the following study, the latent dimension is set to 16. There are several considerations for this choice. First, the dimension of the latent vector should be large enough to maintain adequate complexity; otherwise, the loss of information may lower the performance of the AE or VAE model. Second, although the working condition of the dataset is simplified and defined by 10 independent parameters, it is necessary to maintain the scalability of the network. Finally, an oversized latent vector may result in difficulties in training, and the loss function remains at a relatively high value even when trained excessively. This phenomenon can be considered a kind of training failure brought by the gradient descent algorithm, namely, local optima.

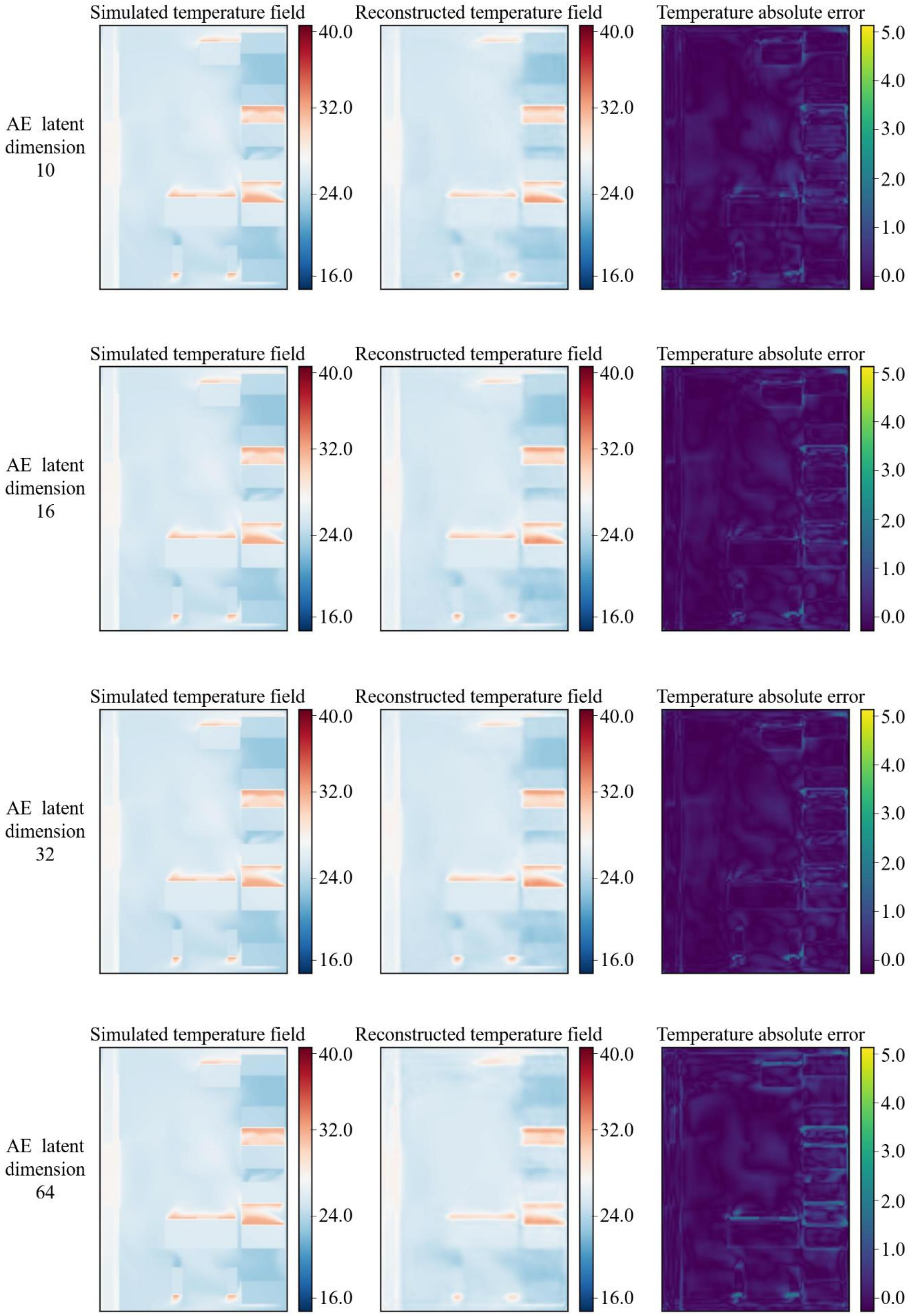
Simulated temperature field
Reconstructed temperature field
Temperature absolute error
AE latent dimension 10
AE latent dimension 16
AE latent dimension 32
AE latent dimension 64
40.0
32.0
24.0
16.0
5.0
4.0
3.0
2.0
1.0
0.0

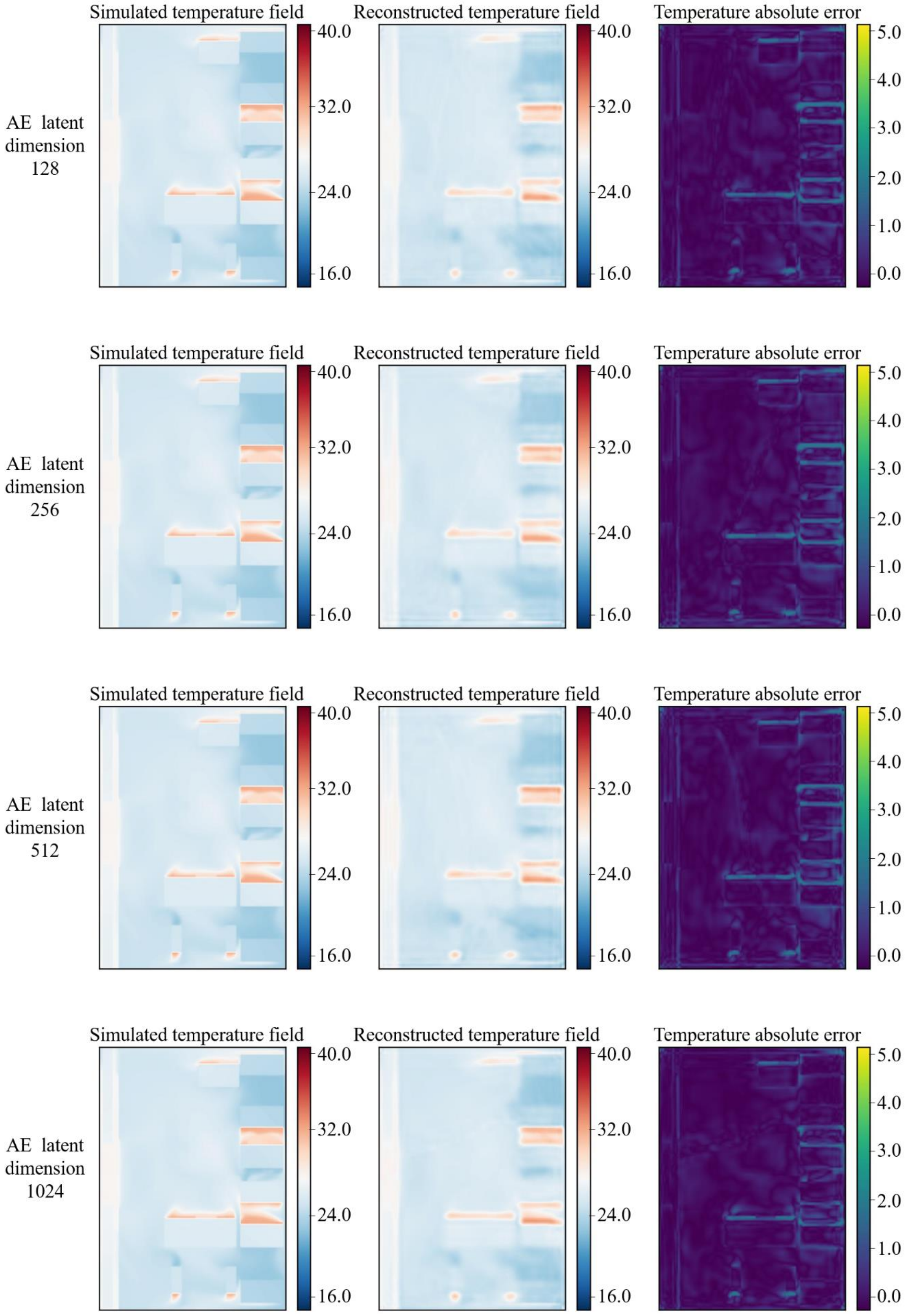

**Fig. 14** Comparison of the AE reconstruction results over the validation set

The structure of the VAE model is established with the model structure and optimal latent dimension by AE. The crux of VAE training is to obtain an optimal mathematical form of the loss function. The weights between MSE and KLD need to be optimized. In this part, the weight parameter (see Eq. 5) is set in the range from $1 \times 10^{-3}$ to $1 \times 10^{-8}$. The comparison of the reconstruction error for different KLD weight parameters is shown in the MAE form, as shown in **Fig. 15**. To demonstrate the performance of VAE models more intuitively, the accuracy of each reconstructed data is calculated and shown in **Fig. 16**, which shows that the accuracies of most results are higher than 90%. By the analysis of the MAE distribution and accuracy distribution, it is concluded that an optimal value of the weight parameter is $1 \times 10^{-5}$, which has a significantly low mean MAE of 0.293℃ and high mean accuracy of 95.58%, and 99.78% of the reconstructed temperature fields have an accuracy over 90%.

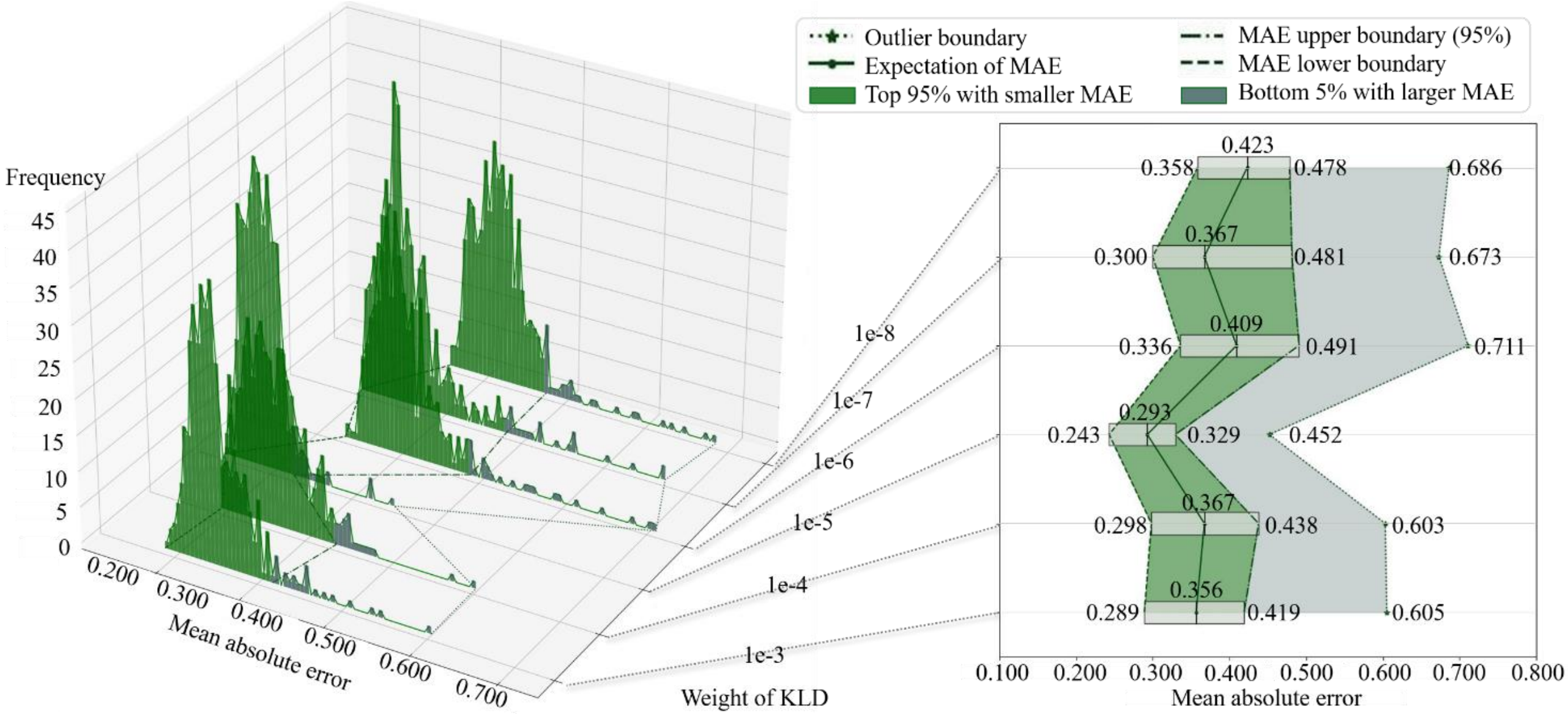

**Fig. 15** Analysis of the mean absolute error of VAE reconstruction over the validation set

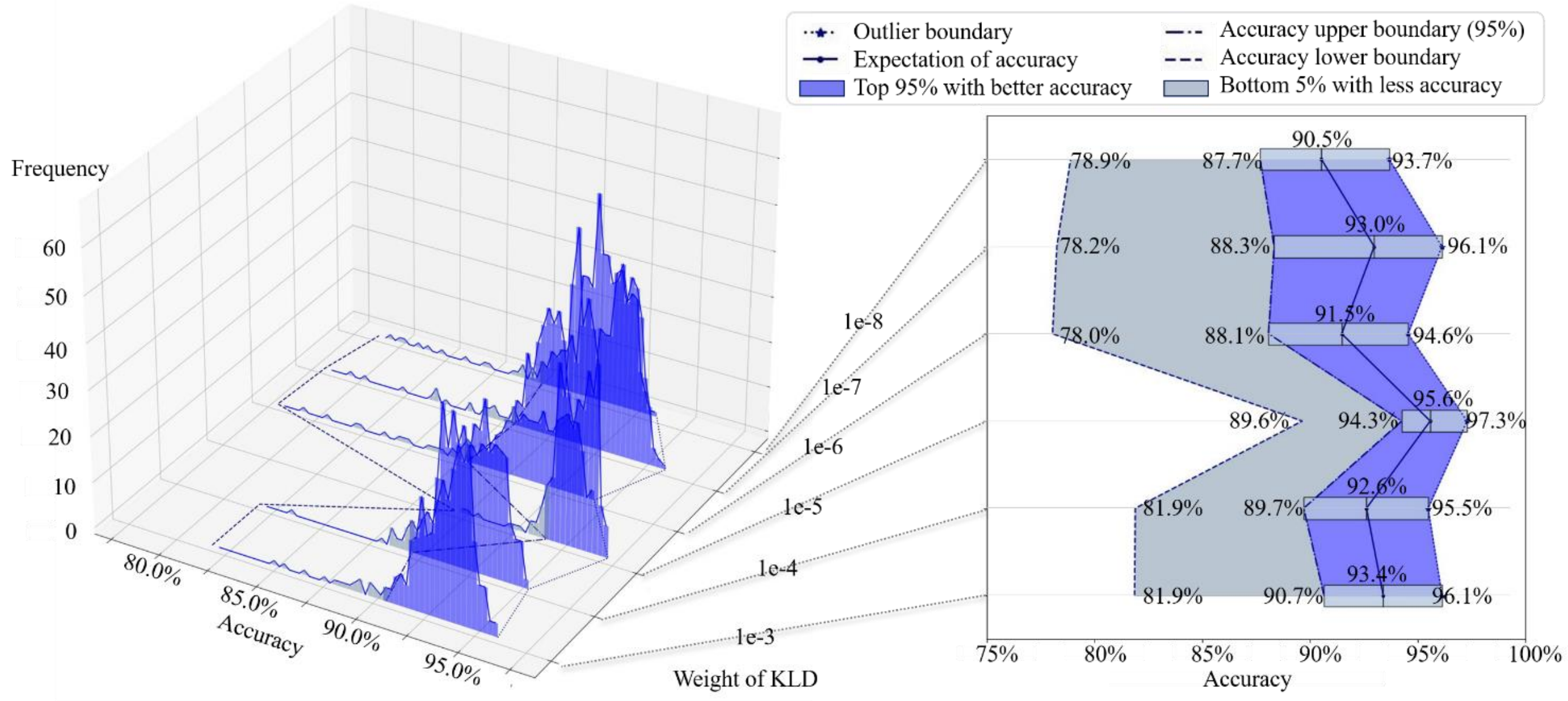

**Fig. 16** Analysis of reconstruction accuracy of VAE reconstruction over the validation set

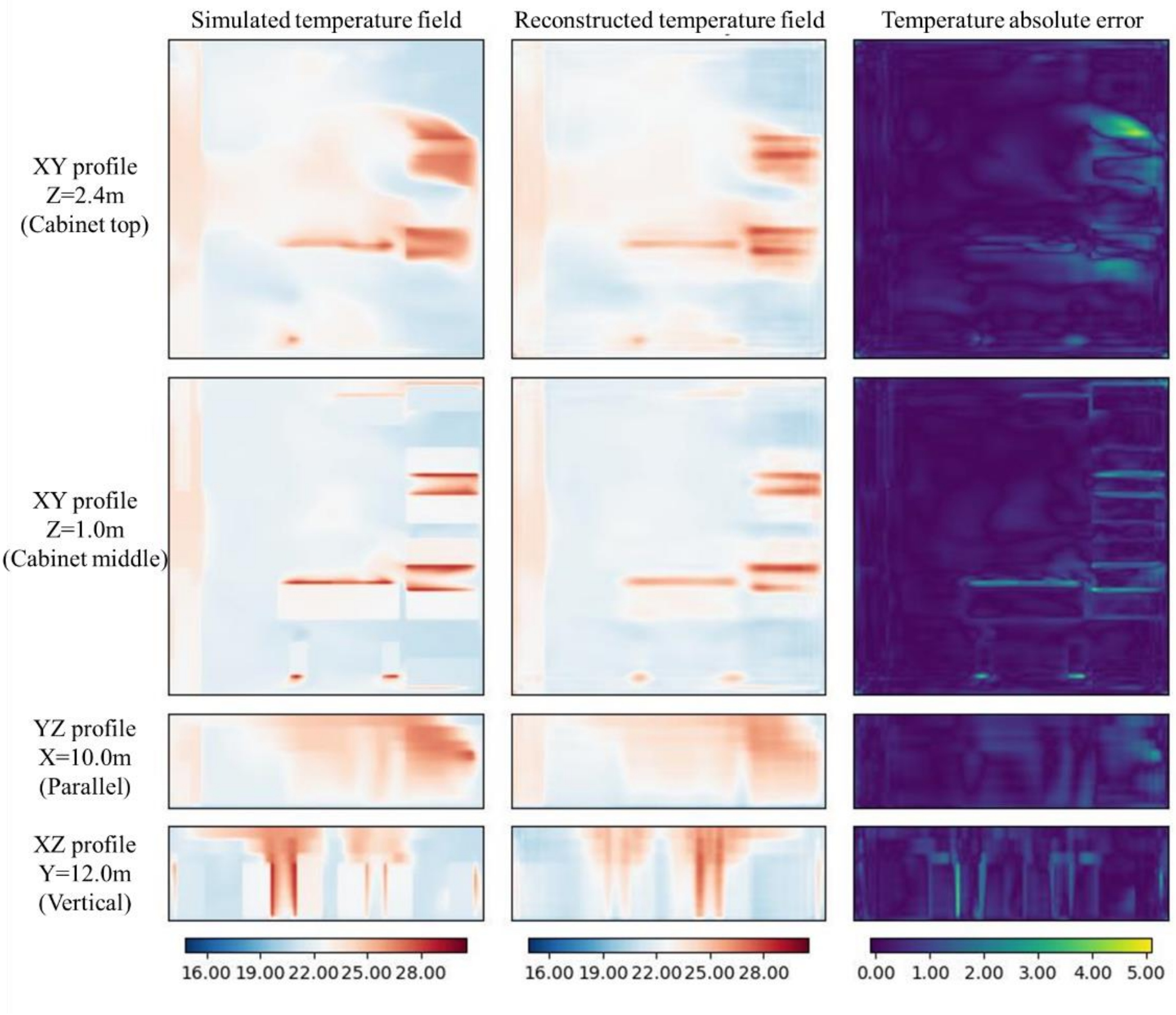

**Fig. 17** VAE reconstruction result over the validation set (best)

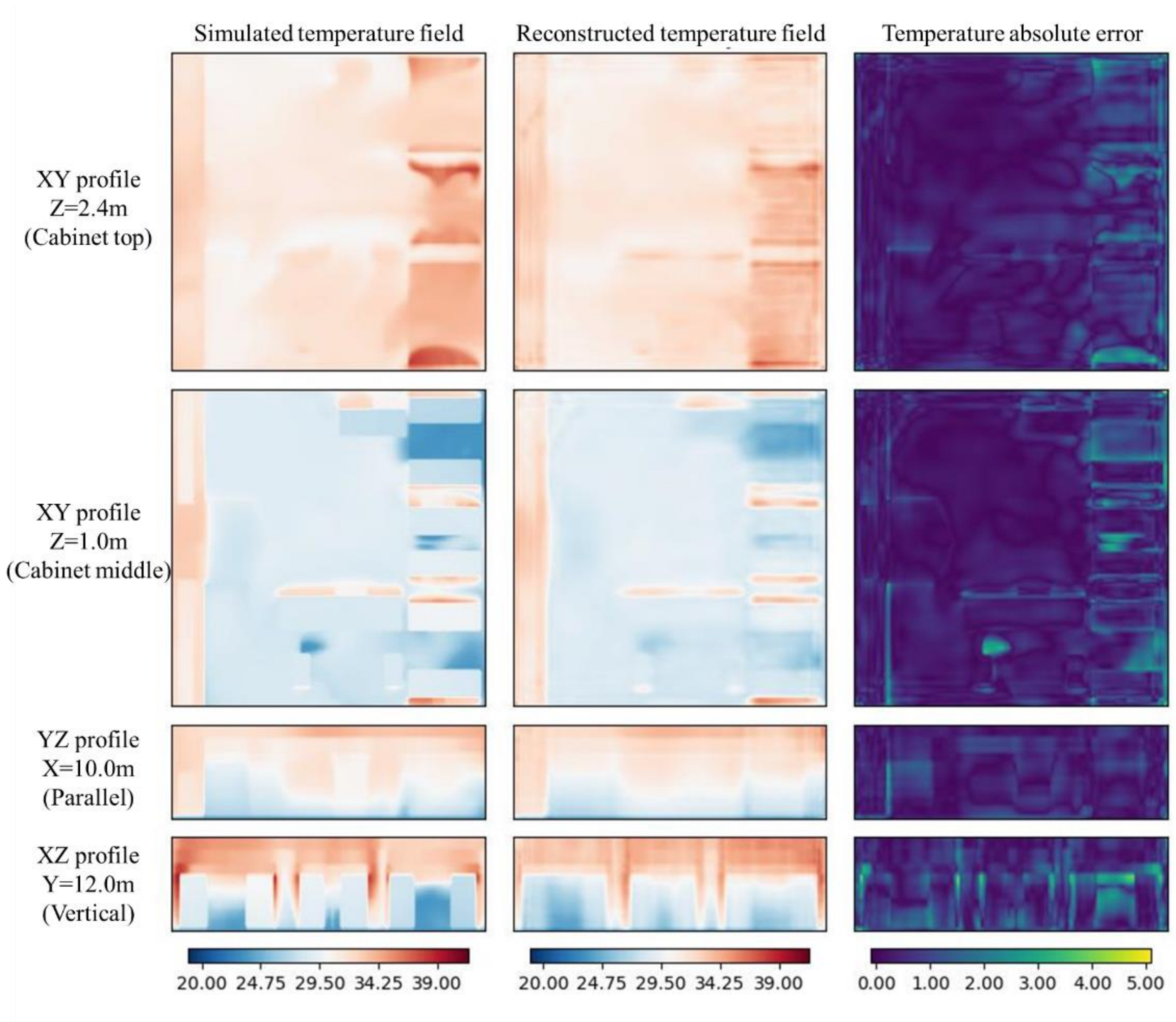

**Fig. 18** VAE reconstruction result over the validation set (worst)

Furthermore, the best and worst samples are selected to demonstrate the reconstruction accuracy. The best constructed sample is shown in **Fig. 17**. The MAE is 0.243 °C, and the accuracy is 97.3%. The reconstructed contour patterns show a good match with the CFD/HT result, and errors are relatively small. The worst sample is shown in **Fig. 18**. The MAE is 0.452 °C, and the accuracy is 89.6%. Although the accuracy is not very high, the reconstructed contour patterns also generally match the CFD/HT result, which demonstrates an acceptable reconstruction performance of the current VAE model.

## B. Performance on scalar physical field prediction

After training, the VAE decoder is then employed to generate a dataset of latent vectors, $z$. Meanwhile, the $z$ set corresponds to the operation parameter dataset, i.e., $p$. An ANN compilation network is established and trained to map the operation parameter $p$ to latent vector $z$. The ANN structure is shown in **Fig. 10**. Compared to VAE, the tuning process of ANN is relatively easy to implement. The best and worst samples precited by the ANN are shown in **Fig. 19**. The abscissa of **Fig. 19** represents the latent vector gained from VAE encoding, and the ordinate represents the latent vector obtained from ANN mapping. For the best sample, all 16 values of latent vector $z$ are mapped correctly and accurately. For the worst sample, most of its values are mapped within an error of $\pm 0.5$ (the gray area). The range of error is relatively small compared with the range of the $z$ value.

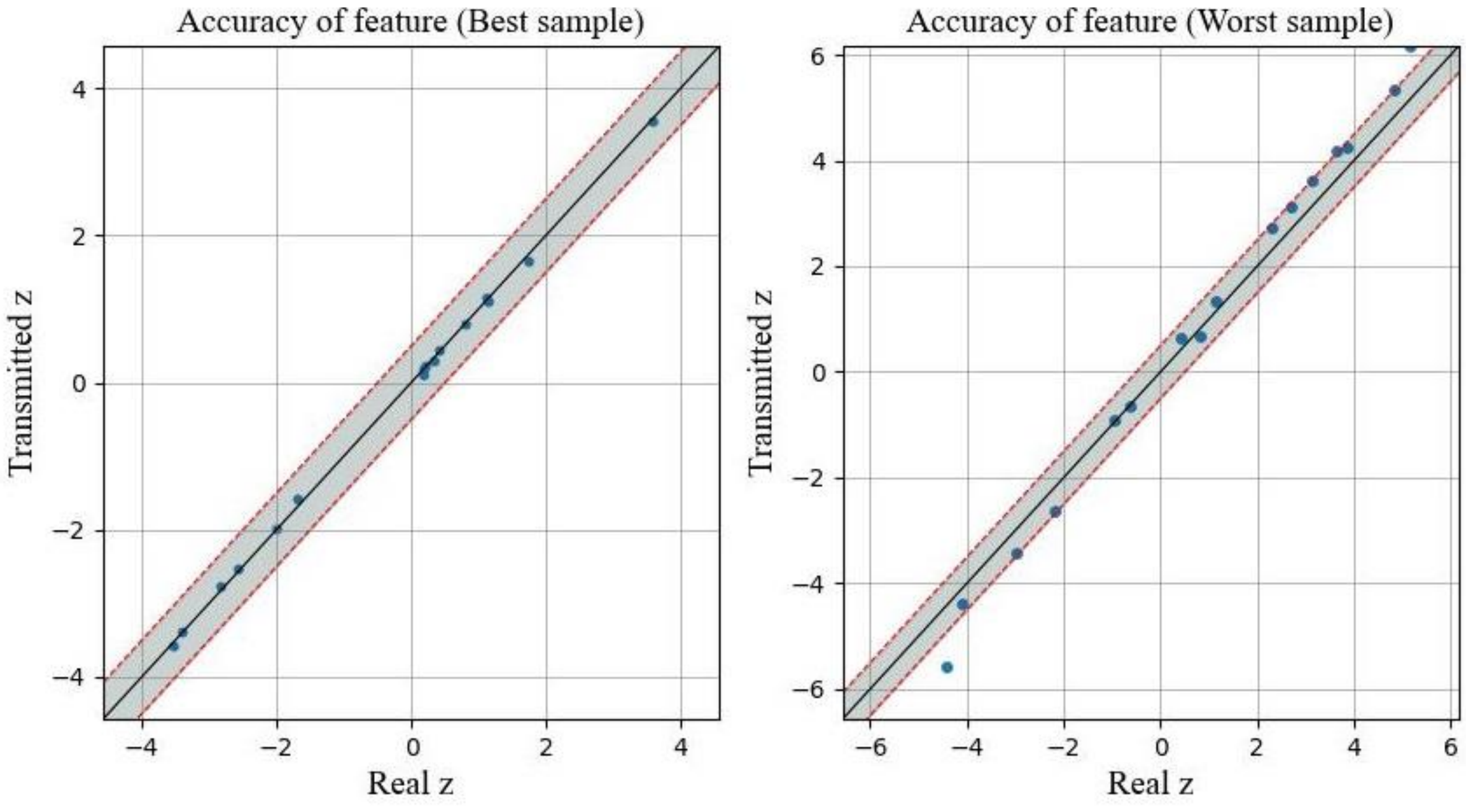

**Fig. 19** Best and worst samples of ANN transition over the test set

Additionally, **Fig. 20** shows comparisons between the encoded and mapped latent vectors for all the samples. The 16 features are compared and shown sequentially. The gray lines mark the equal value for the encoded feature and the mapped feature, and the red lines demonstrate the $\pm 0.5$ error bound.

The scatter plots show that the ANN network is capable of high-accuracy feature compilation because most of the features of all 16 dimensions are compiled accurately. The ranges and mean values of the 16 features are also listed in the lower right table. The distribution shows a concentricity of latent space, which provides an assurance of generalizability.

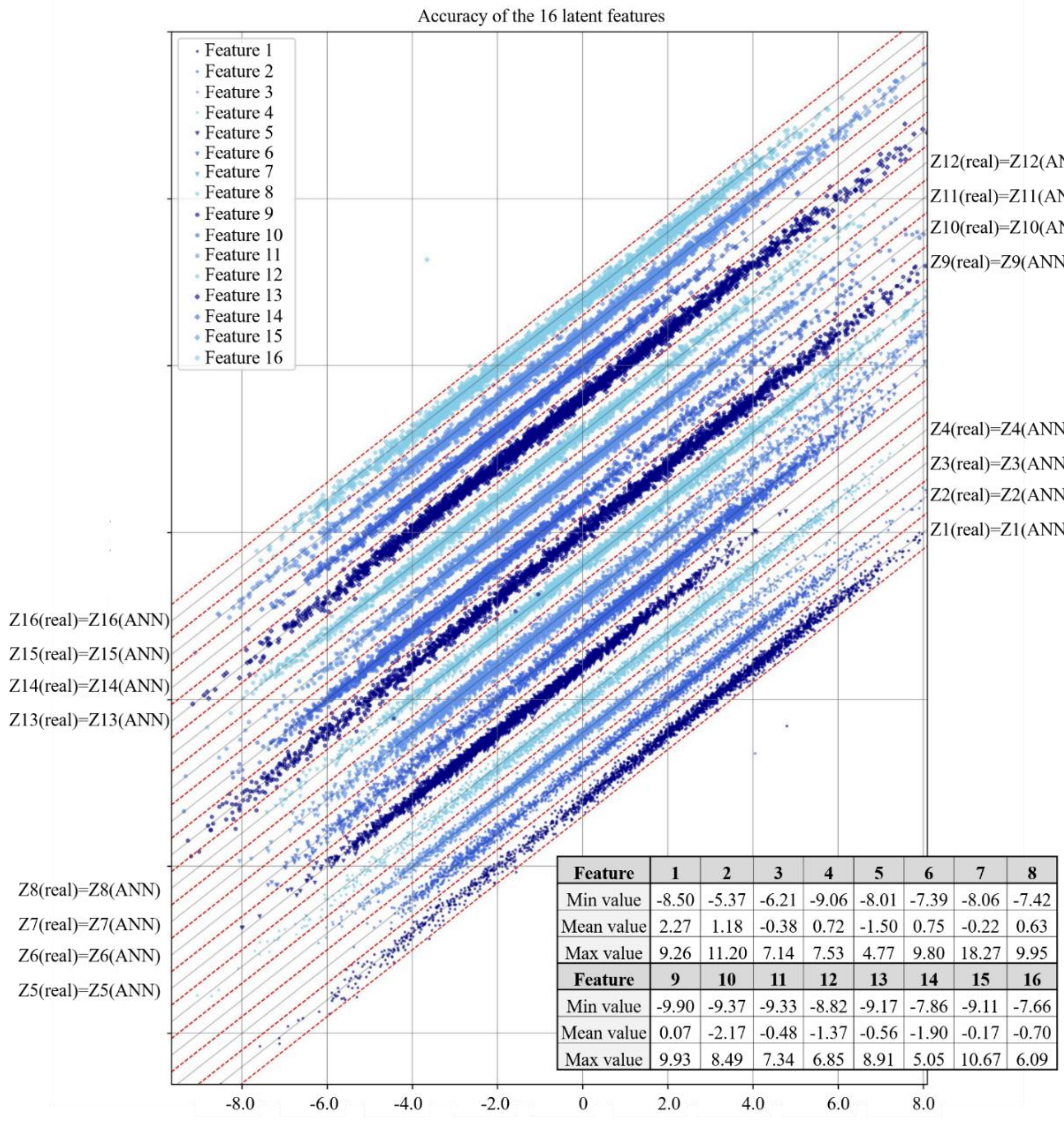

| Feature | 1 | 2 | 3 | 4 | 5 | 6 | 7 | 8 |
|---|---|---|---|---|---|---|---|---|
| Min value | -8.50 | -5.37 | -6.21 | -9.06 | -8.01 | -7.39 | -8.06 | -7.42 |
| Mean value | 2.27 | 1.18 | -0.38 | 0.72 | -1.50 | 0.75 | -0.22 | 0.63 |
| Max value | 9.26 | 11.20 | 7.14 | 7.53 | 4.77 | 9.80 | 18.27 | 9.95 |
| **Feature** | **9** | **10** | **11** | **12** | **13** | **14** | **15** | **16** |
| Min value | -9.90 | -9.37 | -9.33 | -8.82 | -9.17 | -7.86 | -9.11 | -7.66 |
| Mean value | 0.07 | -2.17 | -0.48 | -1.37 | -0.56 | -1.90 | -0.17 | -0.70 |
| Max value | 9.93 | 8.49 | 7.34 | 6.85 | 8.91 | 5.05 | 10.67 | 6.09 |

**Fig. 20** Accuracy of ANN complier over the dataset

Up to this point, the ANN complier and VAE decoder have both been trained and tested separately.

Then, the two models are combined to quickly predict the physical field. The goal is to feed an operation parameter vector $p$ into the ANN-VAE model and have a physical field predicted to be consistent with the CFD/HT simulation result. The test set is fed to the ANN-VAE model. The prediction result is analyzed in terms of MAE and accuracy distributions over the test set, as shown in **Fig. 21.** The mean MAE and mean accuracy over the test set are 0.308℃ and 95.2%, respectively. The temperature distributions of the best prediction sample and the worst prediction are shown in **Fig. 22** and **Fig. 23**, respectively. For the best sample, the MAE is 0.244 °C, and the prediction accuracy compared with the simulation result is 97.3%. It is observed that the predicted temperature field matches the simulated results perfectly. In contrast, the MAE is 0.586℃ for the worst sample, and the accuracy is 89.6%. It is observed that the predicted temperature field also generally matches the simulated results. It is justifiable to conclude that the proposed ANN-VAE prediction model is satisfactory for scalar physical fields.

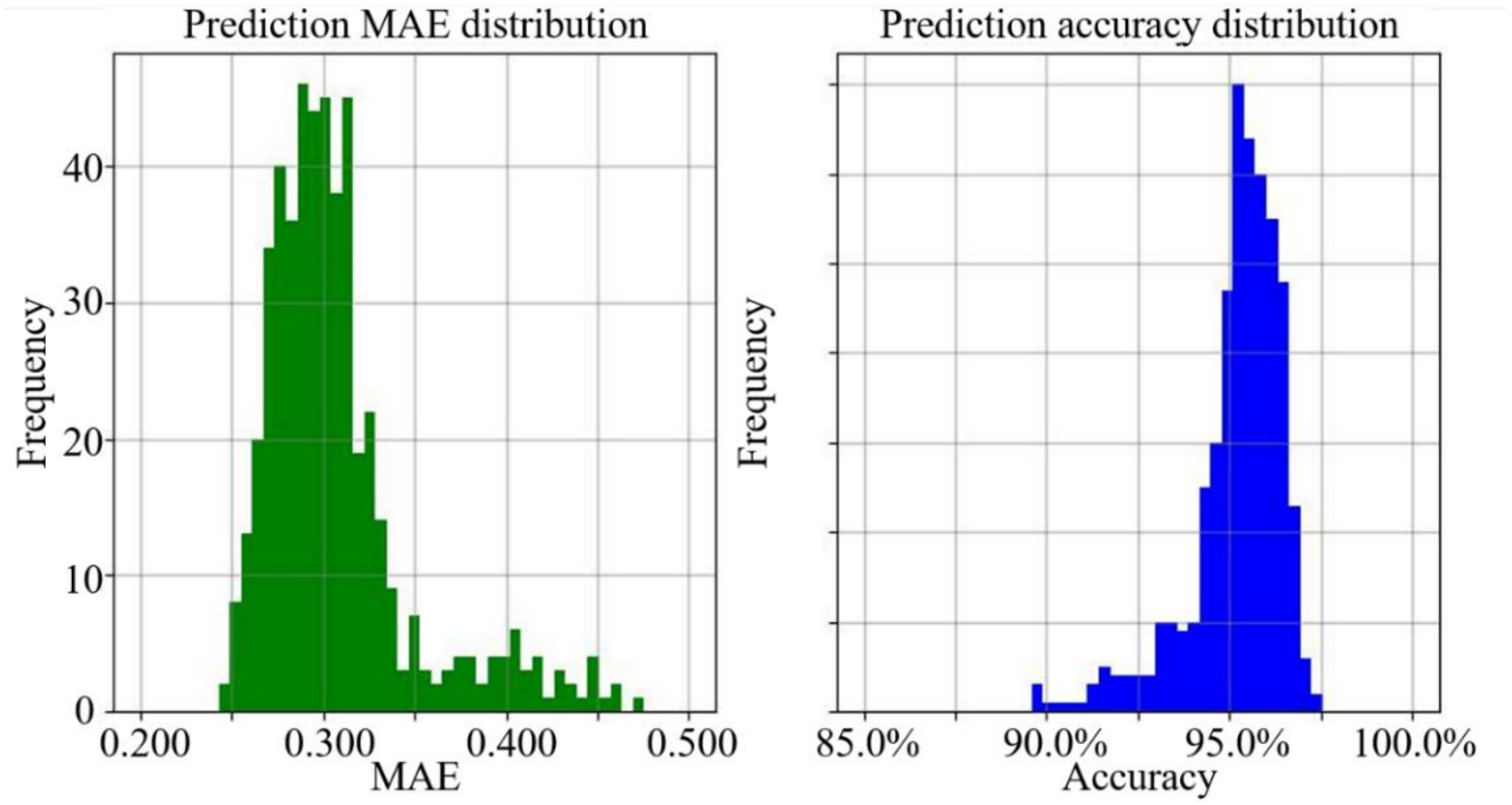

**Fig. 21** MAE and accuracy of ANN-VAE-based prediction over the test set

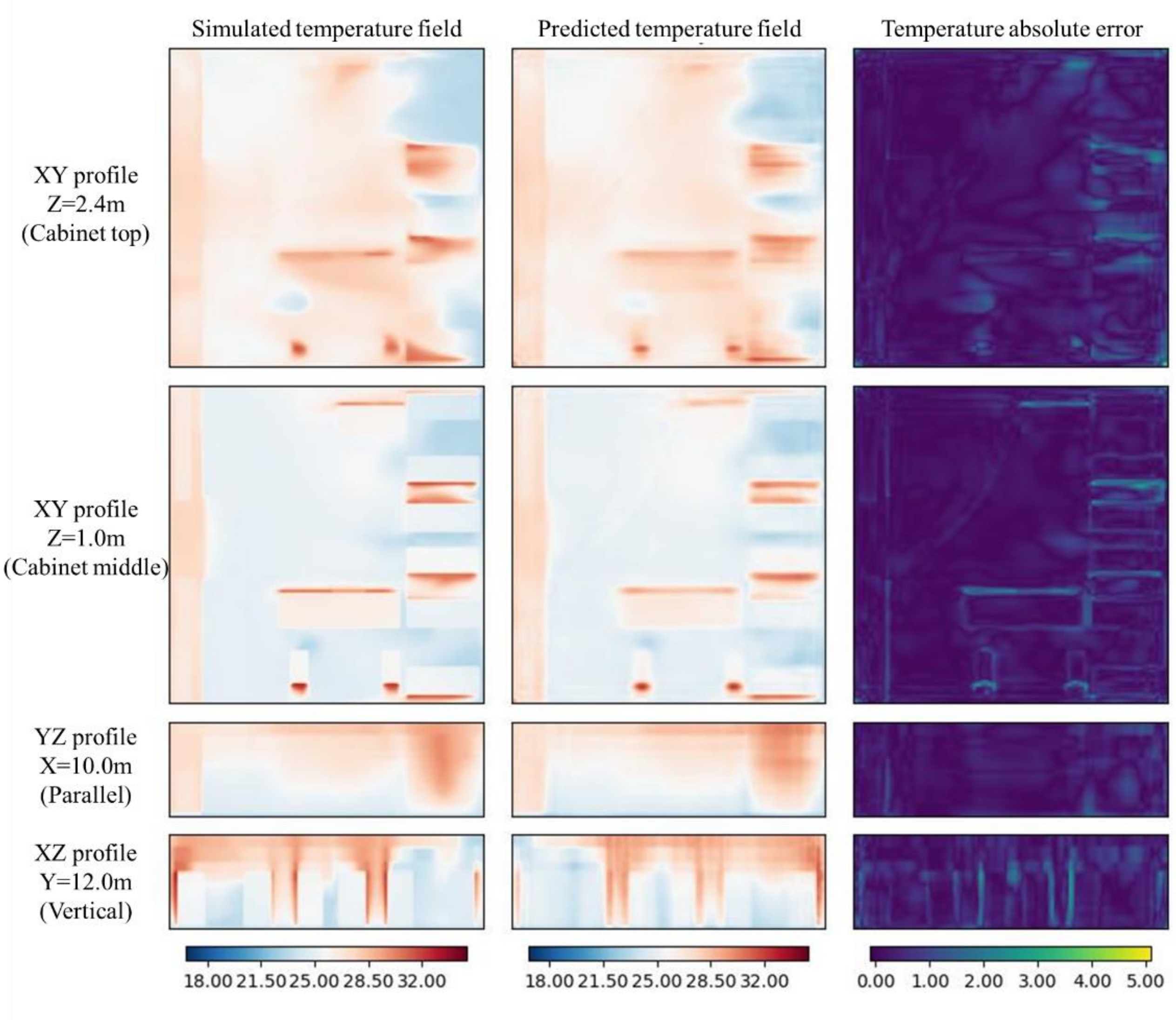

**Fig. 22** Best sample of ANN-VAE prediction over the test set

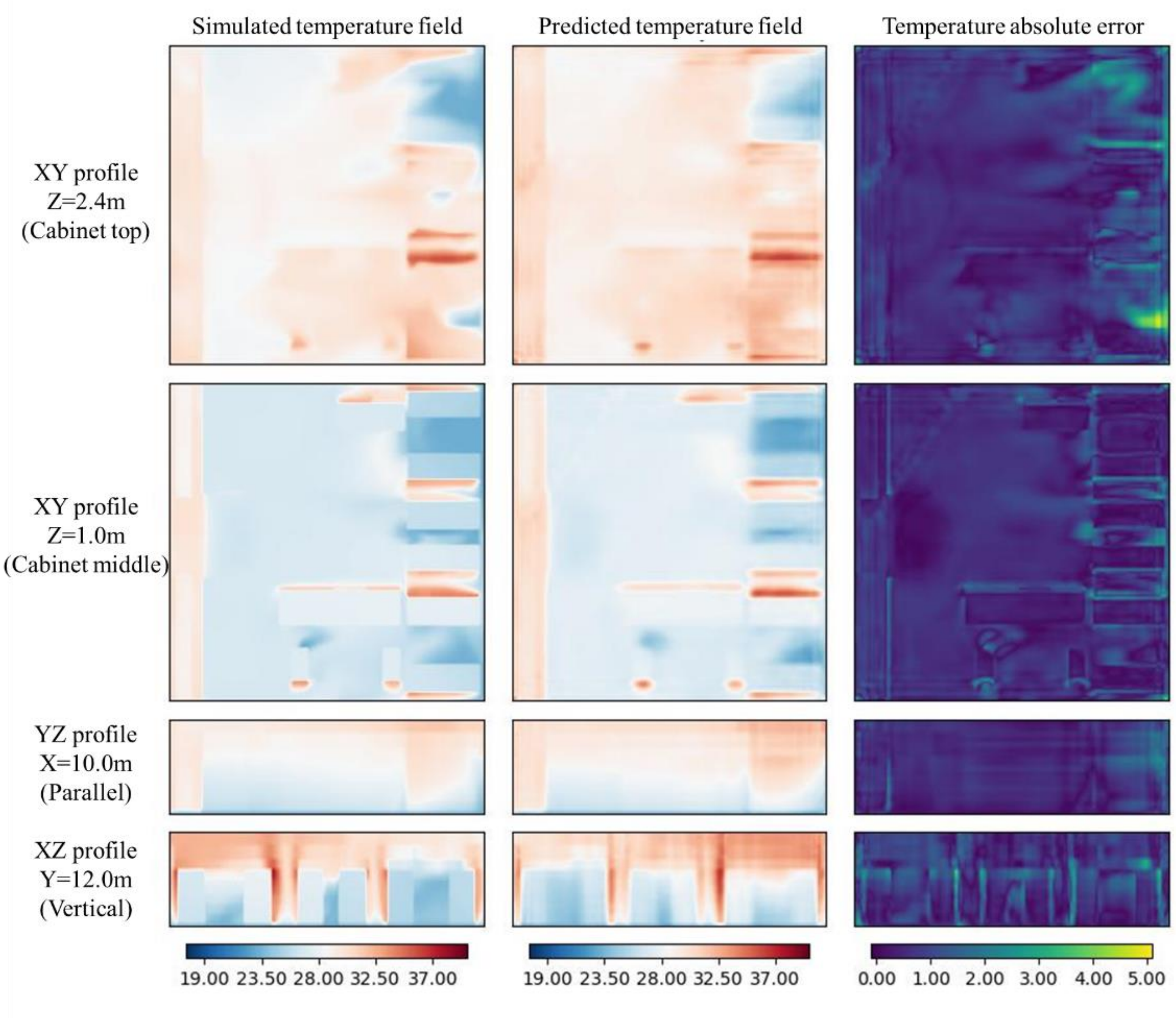

**Fig. 23** Worst sample of ANN-VAE prediction over the test set

## C. Performance on vector physical field prediction

The proposed ANN-VAE model and its training strategy are also effective for the prediction of physical vector fields. Taking the velocity field as an example, the prediction of 3D physical vector fields can be achieved by training each component in sequence. Since each velocity component shares the same size and rank with the temperature, the feature extraction and tuning process are the same. The training processes and error distributions of $u$, $v$ and $w$ are shown sequentially in **Fig. 24**. The detailed prediction performances are listed in **Table 3**. The input velocity ranges from approximately -1 m/s to 1 m/s, and the MAEs of all predictions are below 0.050 m/s. It is observed that most samples are predicted with an accuracy higher than 96%, and very few prediction accuracies are below 90%. This is a valid affirmation for the performance of the ANN-VAE model.

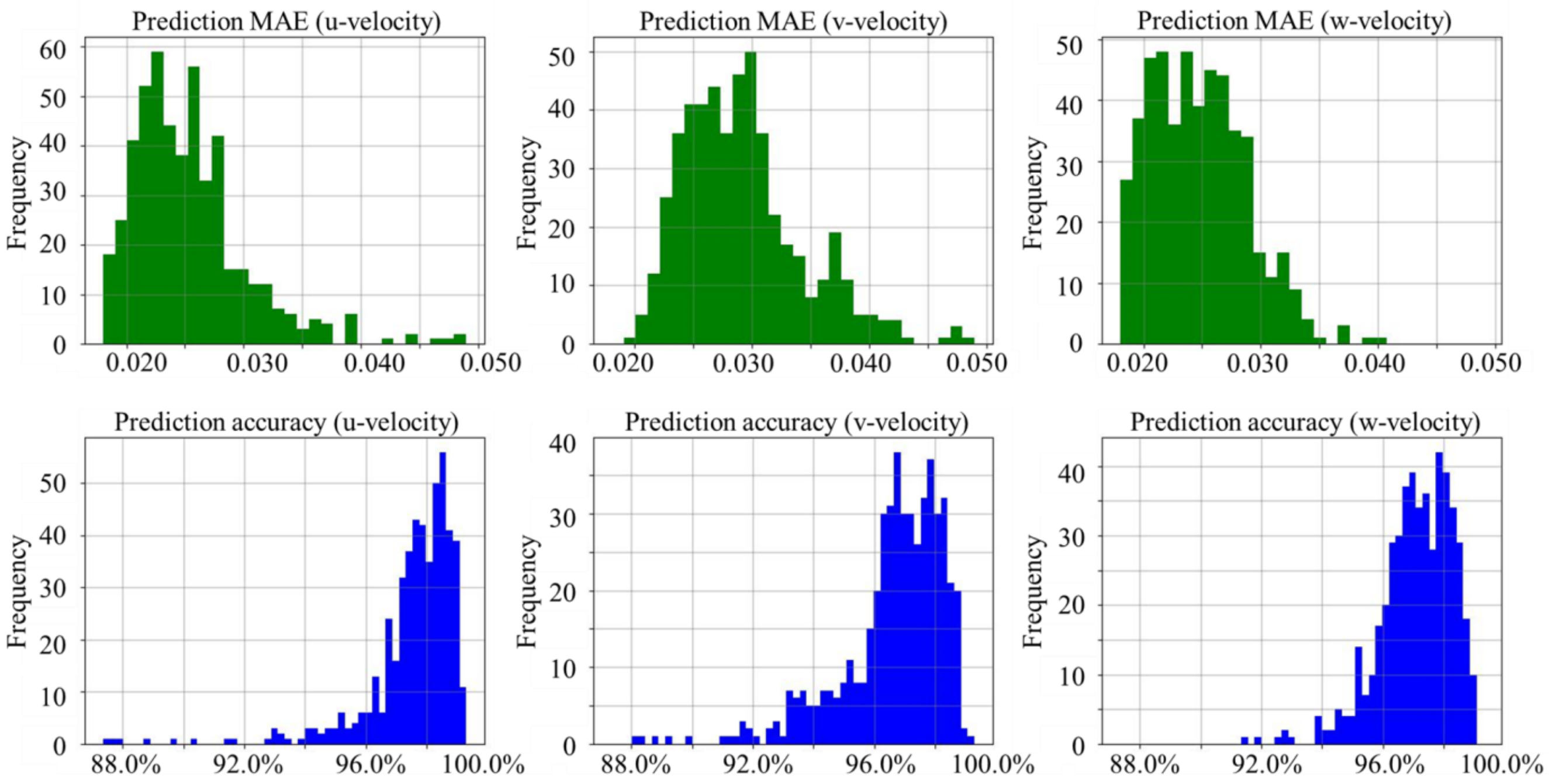

**Fig. 24** MAE and accuracy of ANN-VAE-based prediction over the test set

**Table 3** Performance for velocity fields

| Prediction | Over test set | | At best sample | | At worst sample | |
|---|---|---|---|---|---|---|
| | Mean MAE | Mean accuracy | MAE | Accuracy | MAE | Accuracy |
| u | 0.026 m/s | 97.5% | 0.018 m/s | 99.3% | 0.049 m/s | 87.4% |
| v | 0.029 m/s | 96.6% | 0.020 m/s | 99.1% | 0.049 m/s | 88.1% |
| w | 0.025 m/s | 97.1% | 0.018 m/s | 99.0% | 0.040 m/s | 91.4% |

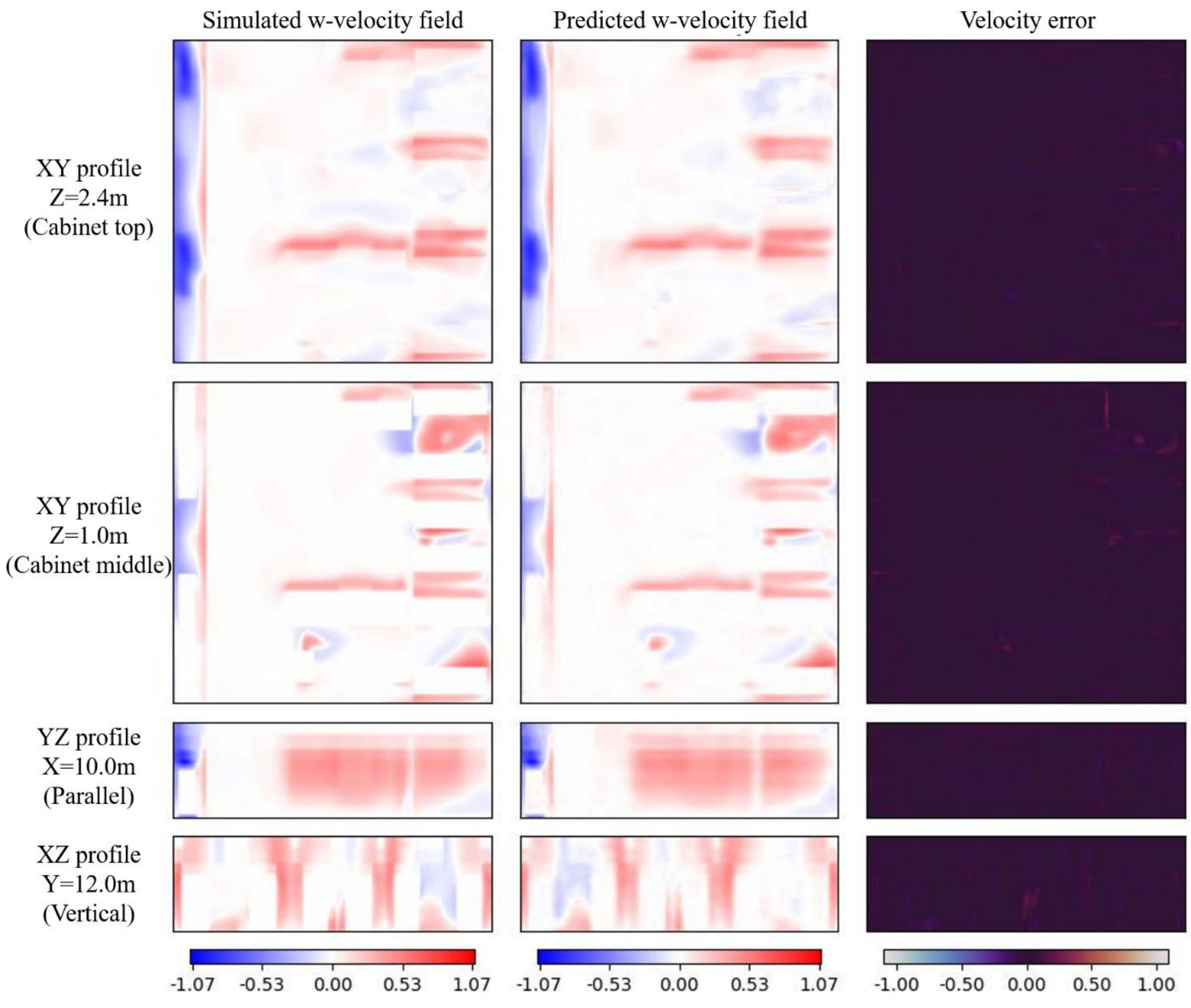

**Fig. 25** Best sample of *w*-velocity prediction over the test set

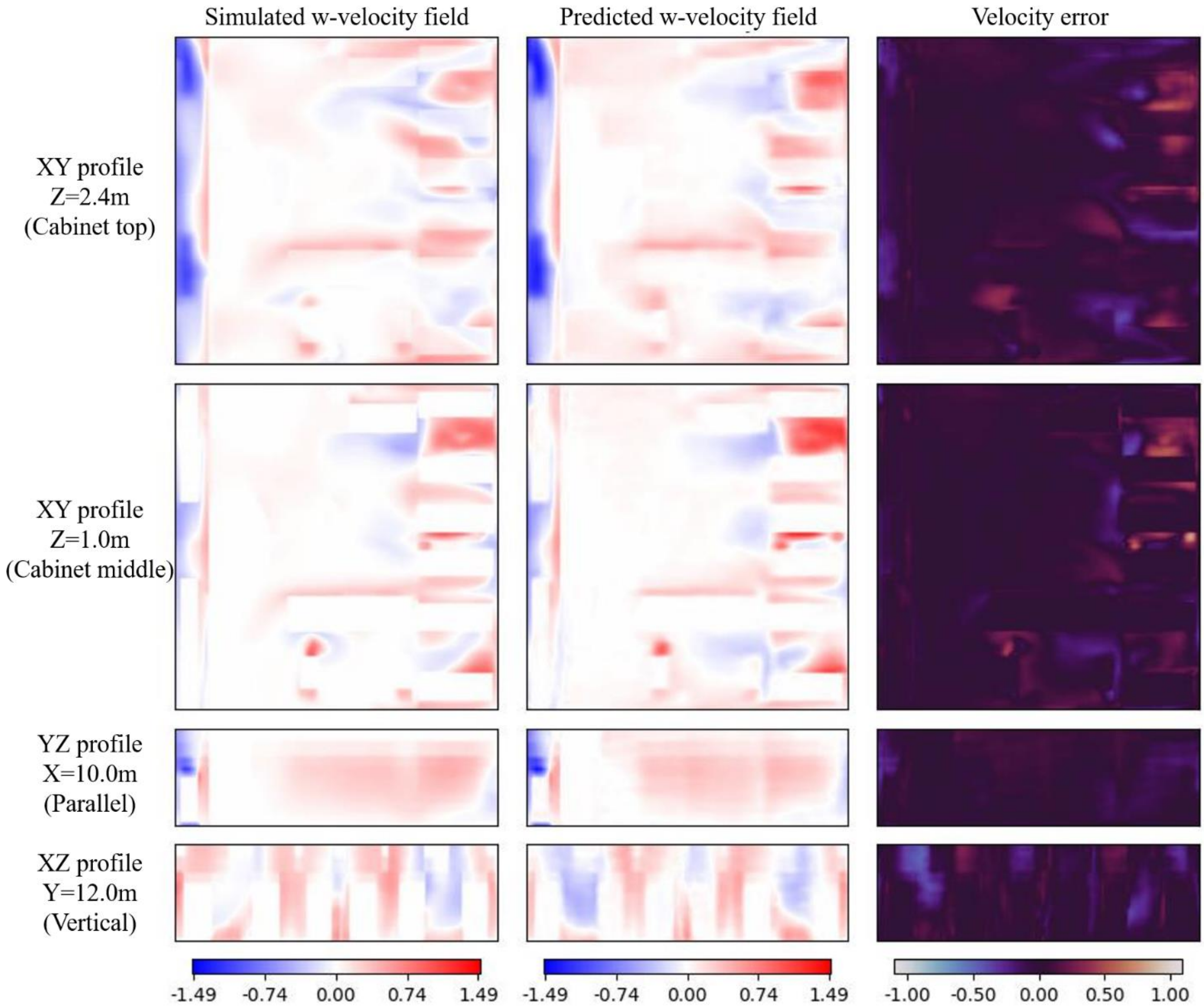

**Fig. 26** Worst sample of *w*-velocity prediction over the test set

In the data center room, the vertical velocity (*w*-velocity) is the dominant velocity component. High accuracy of *w*-velocity prediction guarantees the accuracy of velocity vector field prediction. For *w*-velocity prediction, the mean prediction accuracies are 97.1%. The best sample with the highest accuracy (99.0%) is shown in **Fig. 25**, and the worst sample with the lowest accuracy (91.4%) is shown in **Fig. 26**. For the best samples of *w*-velocity, the contour patterns of the predicted field match well with the CFD/HT simulated field. For the worst samples, the contour patterns also have general consistency with the simulated result.

Prediction for vector velocity fields is achievable with 3 components predicted simultaneously. The magnitude of speed (sqrt($u^2+v^2+w^2$)) is used to evaluate the prediction accuracy. The MAE and accuracy of vector field prediction over the whole dataset are shown in **Fig. 27**, and the detailed results are shown in **Table 4.** A prediction sample of the highest accuracy among the overall dataset is shown in **Fig. 28**. The velocity components parallel (for example, *u*- and *v*-velocities of XY profiles) to the profile are shown as streamlines, and the velocity vertical (for example, *w*-velocity of XY profiles) to the profile is shown as a contour. The absolute errors of velocity magnitude are computed and shown on the right. The worst sample with the lowest accuracy is shown in **Fig. 29**. Although the absolute error of the velocity magnitude is relatively large, it is observed that the predicted airflow organization is barely satisfactory. The general direction and structure of the airflow are predicted successfully. It is concluded that the proposed ANN-VAE model is capable of accurately predicting physical vector fields.

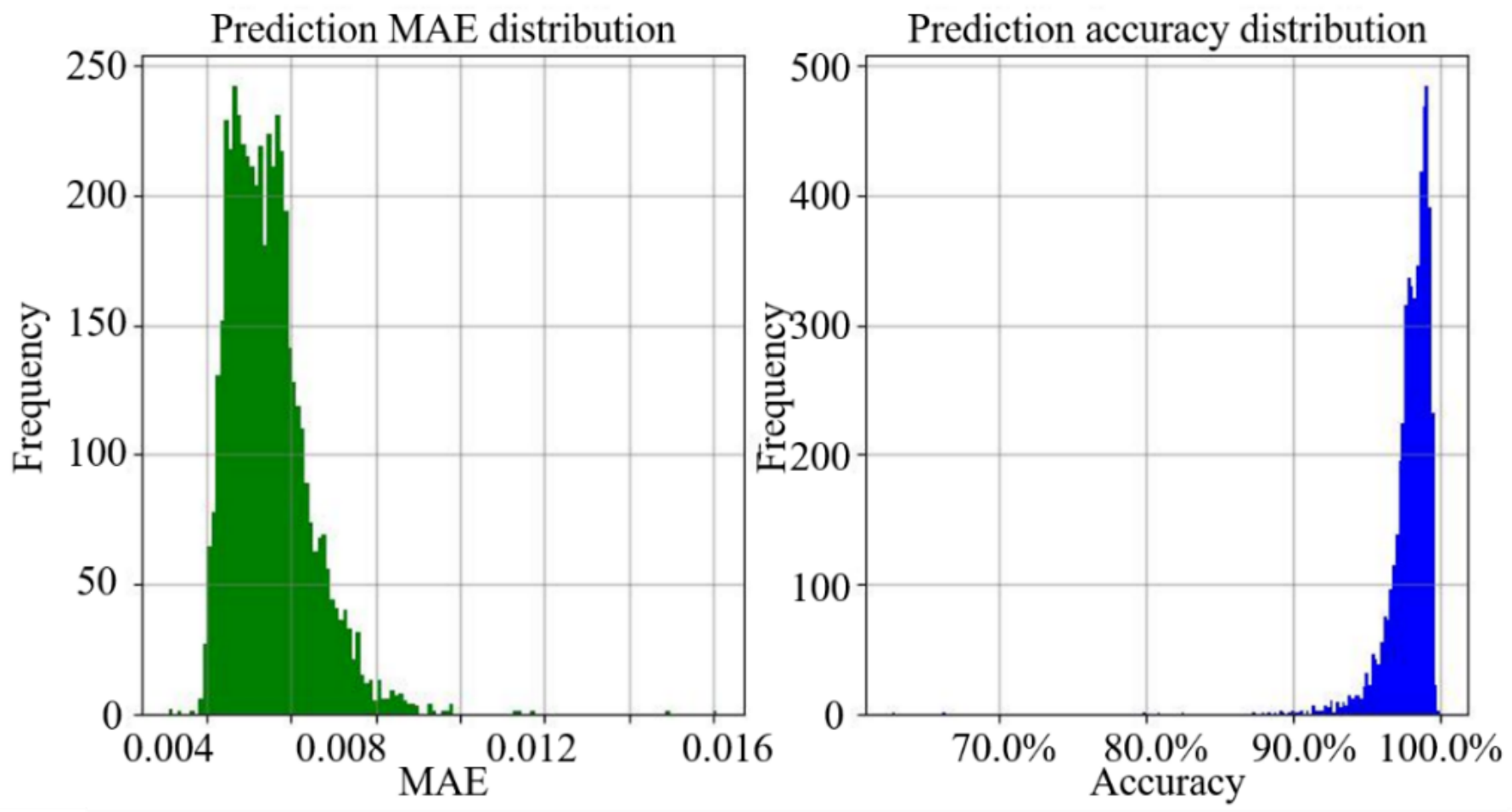

**Fig. 27** Analysis of vector field prediction over the dataset

**Table 4** Performance of vector field prediction over the dataset

| Velocity prediction | MAE | | | Accuracy | | |
|---|---|---|---|---|---|---|
| | Mean | Best sample | Worst sample | Mean | Best sample | Worst sample |
| Velocity | 0.055 | 0.031 | 0.161 | 97.9% | 99.9% | 62.7% |

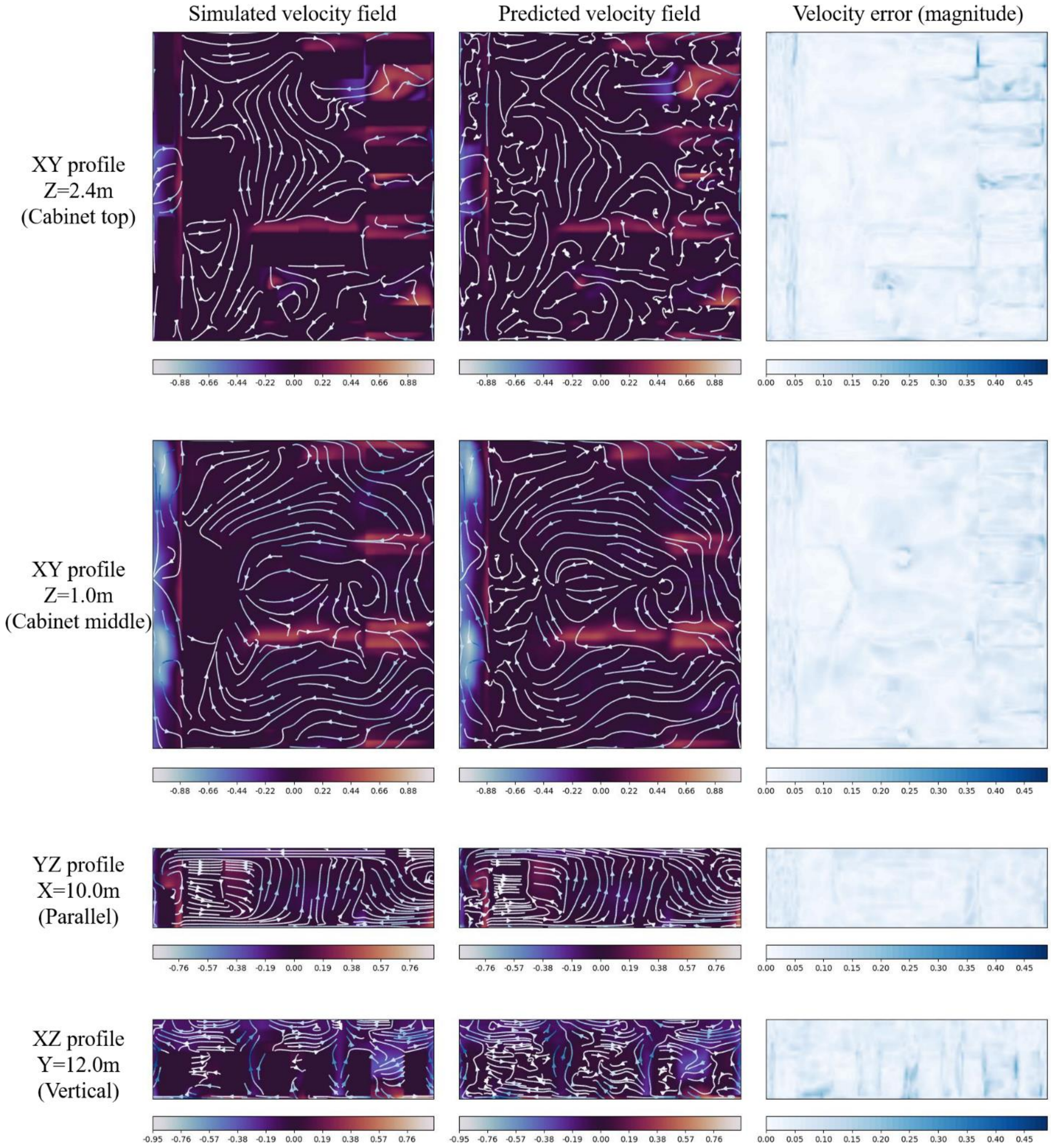

**Fig. 28** 3D vector field prediction (best sample)

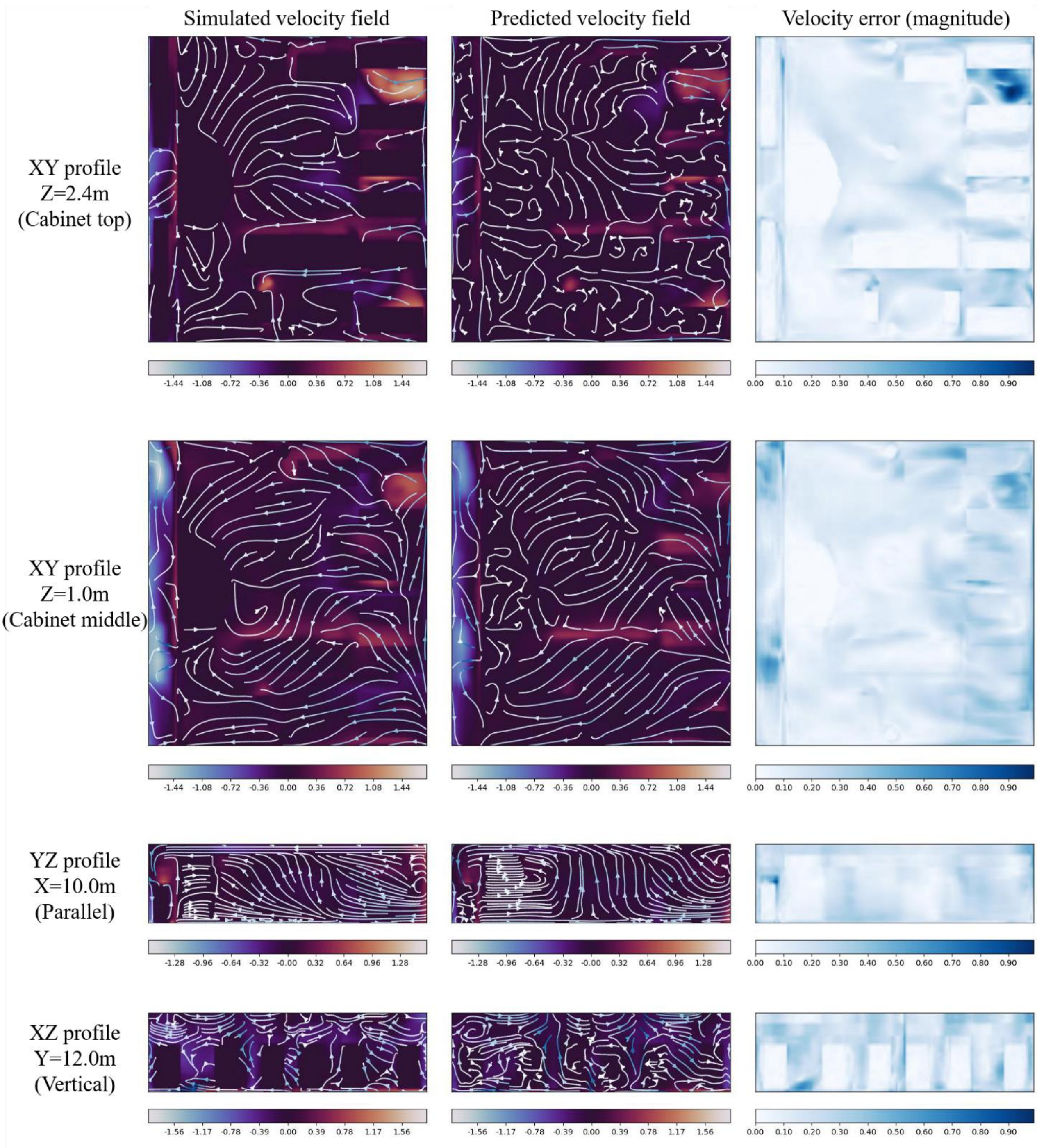

**Fig. 29** 3D vector field prediction (worst sample)

## V. CONCLUSION

This paper introduced a novel method for 3D full-field prediction based on an ANN-VAE composite model. The method is experimentally demonstrated to be capable of accurately predicting steady scalar and vector fields of a certain engineering problem. The work can be summarized as follows:

(1) A VAE generative model is proposed with a detailed training strategy for flow field reconstruction. First, the structure of VAE is ascertained by a data compression AE network. Then, the VAE network is trained and applied to the reconstruction of physical fields for the data center cooling cycle.

(2) An ANN-VAE prediction model is established and trained based on the acquired VAE model. The ANN model is employed to map the 10 operation parameters to latent vectors with 16 values that can be recognized by the VAE decoder. Then, the decoder of the VAE is employed to reconstruct the latent vector into predicted physical fields.

(3) The overall methodology proposed is expected to be functional for predictions of both physical scalar and vector fields. In this paper, fields of size $137 \times 131 \times 20$ are predicted with relatively high accuracy and a significantly small time cost, as listed in **Table 5**.

**Table 5** CPU time cost of physical field computation

| CPU time cost | Temperature | u-velocity | v-velocity | w-velocity | Overall |
|---|---|---|---|---|---|
| CFD/HT method | | | 1 h:39 min:35s = 5975s | | |
| ANN-VAE method | 0.01563s | 0. 01563s | 0. 01563s | 0. 01563s | 0.06252s |

With the advancement of AI technology, it is generally believed that new light will be revolutionarily shed onto various topics. In this paper, we mainly realize the runtime prediction of 3D physical fields in data centers under steady state conditions, and there is still much room for the expansion of related research methods, which are expected to be applied to more complex problems, such as periodic changes in temperature fields and turbulent convective heat transfer.

In subsequent research, the ANN-VAE network built in this paper can be further adapted for generation tasks of various physical fields, such as temperature, pressure and other types of flow fields. Furthermore, the application of the research content of this paper is general, and the method can be further extended to many other studies, such as aviation, aerospace, fluid mechanics, convective heat transfer and building environments.

**ACKNOWLEDGMENTS**

This work was supported by the National Natural Science Foundation of China (11872230, 92152301).